\documentclass[wocolumn, preprint2]{aastex631}

\usepackage{amsmath}

\usepackage{fancybox}

\newcommand{\shadowpicture}[1]{%
  \shadowsize=1pt
  \fboxrule=1pt
  \fboxsep=0pt
  \color{gray}
  \shadowbox{\fboxsep=3pt\fcolorbox{white}{white}{#1}}
  \normalcolor
}

\pdfoutput=1

\received{July 22, 2024}

\revised{September 11, 2024}

\accepted{September 19, 2024}

\submitjournal{ApJ}

\shorttitle{The Spitzer Background ISM}

\shortauthors{Boersma~et~al.}

\begin{document}

\title{The Background Interstellar Medium as Observed from Off-Order Low-Resolution Spitzer-IRS Spectra}

\correspondingauthor{C.~Boersma}
\email{Christiaan.Boersma@nasa.gov}

\author[0000-0002-4836-217X]{C.~Boersma}
\affiliation{NASA Ames Research Center, MS 245-6, Moffett Field, CA 94035-1000, USA}

\author[0000-0002-1440-5362]{J.D.~Bregman}
\affiliation{NASA Ames Research Center, MS 245-6, Moffett Field, CA 94035-1000, USA}

\author[0000-0002-6049-4079]{L.J.~Allamandola}
\affiliation{NASA Ames Research Center, MS 245-6, Moffett Field, CA 94035-1000, USA}

\author[0000-0002-8341-342X]{P.~Temi}
\affiliation{NASA Ames Research Center, MS 245-6, Moffett Field, CA 94035-1000, USA}

\author[0000-0003-2552-3871]{A.~Maragkoudakis}
\affiliation{NASA Ames Research Center, MS 245-6, Moffett Field, CA 94035-1000, USA}

\begin{abstract}

Spitzer `hidden' observations of the background are used to construct a catalog of 4,090 spectra and examine the signature of polycyclic aromatic hydrocarbon (PAH) molecules and their connection to extinction by dust.

A strong positive correlation is recovered between WISE12, E(B-V), and the 11.2 µm PAH band. For  0.06 $\leq$ E(B-V) $\leq$ 5.0, correlations of the 6.2, 11.2, and 12.7 {\textmu}m PAH band are positive with E(B-V). Three dust temperature regimes are revealed. Correlations with WISE12 are well-constrained and that with 12.7/11.2 is flat.

Decomposition with the NASA Ames PAH IR Spectroscopic Database reveals a tentative positive correlation between the 6.2/11.2 and the PAH ionization fraction, while that with 12.7/11.2 is slightly negative, suggesting PAH structural changes. The relation with PAH size and 6.2/11.2 is negative, while that with 12.7/11.2 is positive.

Averaging spectra into five E(B-V) and three T$_{\rm dust}$ bins shows an evolution in PAH emission and variations in 12.7/11.2. Database-fits show an increase in $f_{\rm i}$ and the PAH ionization parameter $\gamma$, but a more stable large PAH fraction. While the largest $\gamma$s are associated with the highest T$_{\rm dust}$, there is no one-to-one correlation. The analysis is hampered by low-quality data at short wavelengths.

There are indications that PAHs in the more-diffuse backgrounds behave differently from those in the general interstellar medium. However, they are often still associated with larger scale filamentary cloud-like structures.

The spectra and auxiliary data have been made available through the Ames Background Interstellar Medium Spectral Catalog and may guide JWST programs.

\end{abstract}


\section{Introduction}
\label{sec:introduction}

The life cycle of interstellar polycyclic aromatic hydrocarbon (PAH) molecules starts with their formation in the ejecta from carbon-rich AGB stars followed by a residency in the interstellar medium (ISM), where they subsequently evolve via processing by ultraviolet (UV) photons. After some 10$^{\rm 7}$~yr, PAHs are incorporated into dark clouds where they are thought to freeze out onto dust grains and are processed further,  along with new PAH formation, now via ice grain chemistry. Once new stars form, PAHs are exposed to UV radiation and processed yet again. This PAH-evolution is observed as spectral changes from one environment to another and within individual objects, e.g., in reflection nebulae (RNe) as a function of distance from the illuminating star \citep[see e.g.,][]{2005ApJ...621..831B, 2016ApJ...832...51B}. PAHs have been well-studied in planetary nebulae (PNe), H~II-regions, and RNe, but little work has been done on PAHs in the background, diffuse ISM.

The direct association of dust with PAHs has been inferred from the correlation of IRAS 100~{\textmu}m measurements with that of the 3.3~{\textmu}m PAH band strength \citep[][]{1996PASJ...48L..53T} and with 4.5-11.7~{\textmu}m spectra \citep[][]{1996PASJ...48L..59O} obtained by the Infrared Telescope in Space (IRTS). While the 3.3~{\textmu}m data show a correlation with both IRAS 12~{\textmu}m (PAHs) and the 100~{\textmu}m (classical dust) data in the intensity range corresponding to the diffuse background ISM, almost all of the mid-IR data sample directions towards dense clouds rather than the background diffuse ISM. Broadband WISE 12~{\textmu}m band 3 data was selected to sample PAH emission based on the IRTS results. Consequently, WISE band 3 data is commonly used as a proxy for PAH emission \citep[see e.g.,][]{2014ApJ...781....5M, 2015MNRAS.452.3629L}.

Utilizing data obtained by the InfraRed Spectrograph \citep[IRS;][]{2004ApJS..154...18H} onboard the Spitzer Space Telescope \citep[Spitzer;][]{2004ApJS..154....1W} during its cryogenic mission that ended on May 15, 2009, this work generates a catalog of 4,090 low-resolution spectra of the background ISM by extracting them from the off-order positions. In turn, this catalog is used to study the PAH emission from the background ISM and its connection to other dust components. The catalog is made available as The Ames Background ISM Spectral Catalog and can be accessed through a comprehensive website.

This paper is organized as follows. Section~\ref{sec:observations} describes the observations and data reduction, Section~\ref{sec:analysis} sets out the analysis, Section~\ref{sec:discussion} discusses the results, Section~\ref{sec:implications} draws astronomical implications, Section~\ref{sec:catalogue} describes The Ames Background ISM Spectral Catalog, and Section~\ref{sec:conclusions} concludes the paper with a summary and its main takeaways.

\section{Observations}
\label{sec:observations}

\subsection{Spitzer}
\label{subsec:spitzer}

\begin{figure*}
  \centering
  \includegraphics[width=\linewidth]{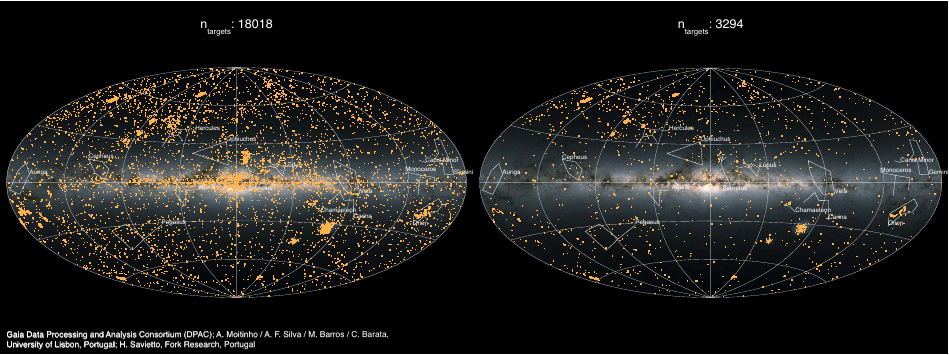}
  \caption{Positions of the Spitzer-IRS staring observations considered in this work overlain on a GAIA all-sky color image using an isotropic Aitoff projection. Some well-known constellations are indicated. Left: All 18,018 irsstare observations. Right: The 3,294 irsstare observations with 60 or 240~s ramp time and outside one-degree of the Galactic plane. See Sect.~\ref{sec:observations} for details.}
  \label{fig:positions}
\end{figure*}

Candidate Spitzer observations were taken from The Nominal Science Operations - Schedule of Executed Science and Calibration Observations, which was obtained from IRSA\footnote{\url{irsa.ipac.caltech.edu/data/SPITZER/docs/files/spitzer/spitzer_obslog.txt}}. The observation log lists 58,575 entries up to the end of the cryogenic mission, of which 18,018 are labeled as \texttt{AOT=irsstare}. The left panel of Fig.~\ref{fig:positions} shows the positions of these observations on the sky overlain on an all-sky GAIA color image (2$^{\rm nd}$ data release\footnote{\url{sci.esa.int/gaia/60196-gaia-s-sky-in-colour-equirectangular-projection/}}). To avoid the crowded Milky Way, a one-degree exclusion zone around the galactic plane is imposed, i.e., half a degree above and below it, based on the reported target positions in the observation log. This brings the number of IRS staring observations down to 17,600. Next, meta data associated with each observation were retrieved using the API\footnote{\url{sha.ipac.caltech.edu/applications/Spitzer/SHA/help/doc/api.html}} exposed by the Spitzer Heritage Archive (SHA) using the AOR key associated with each observation. Subsequently, these were used to check for Short-Low (SL) observations and ensure both SL1 ($7.5\lesssim\lambda\lesssim14.5$~{\textmu}m) and SL2 ($5.2\lesssim\lambda\lesssim14.5$~{\textmu}m) orders are present. This brings the number of observations to consider down to 10,190. Lastly, using the same meta data from the SHA the observations were further limited to those having either a ramp time of 60 or 240~s to ensure enough exposure to detect the weak emission associated with the background ISM. This reduced the number of observations to 3,294; with 2,813 and 481 having 60 and 240~s ramp times, respectively. The galactic positions of these observations are shown in the right panel of Fig.\ref{fig:positions}.

\subsection{Data Reduction}
\label{subsec:reduction}

The raw data associated with each of the 3,294 observations were retrieved from the SHA\footnote{\url{sha.ipac.caltech.edu/applications/Spitzer/SHA/}}. The CUPID software tool at version 2.0\footnote{\url{irsa.ipac.caltech.edu/data/SPITZER/docs/dataanalysistools/tools/cupid/}} was used to generate Basic Calibrated Data (BCD), taking into account different pointings-—traced by the \texttt{CLNUMPOS} FITS header keyword—-in a single observation. BCDs were created both with and without using the pipeline's default dark subtraction. BCDs for a total of 4,090 positions were generated, where those having multiple pointings and those only containing peak-up data were taken into account. It is noted that for a few pointings the slit information was missing (\texttt{PA\_SLIT} FITS header keyword). Table~\ref{app:tab:complications} in the Appendix lists the five observations that could not be processed.

\begin{figure*}
  \centering
  \includegraphics[width=\linewidth]{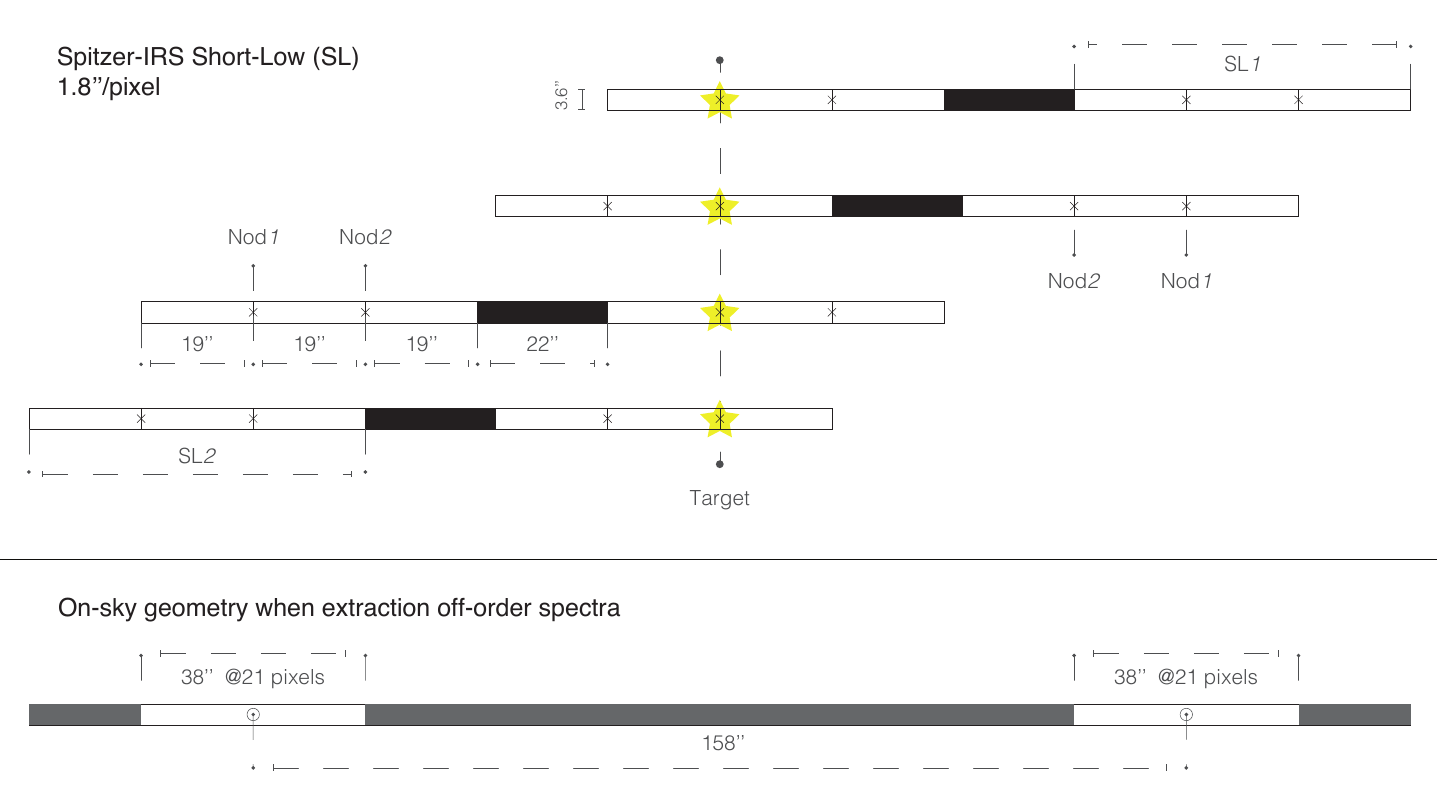}
  \caption{Slit layout. The top four narrow rectangles resemble the long slit of the low-resolution (SL) Infrared Spectrograph (IRS) module onboard Spitzer projected onto the sky. Each slit depicts the position of the target (yellow star) in the slit when employing a nodding strategy. When the target is, in either nod position, on the left half of the slit its light is diffracted by the grating in first order (SL1). Subsequently, when the source is on the right half the light is diffracted in second order (SL2). As a bonus, the SL2 configuration also produces a spectral segment covering 7.3 $\lesssim\lambda\lesssim$ 8.7~{\textmu}m (SL3). Note that the 22" part of the slit shaded black is not used. Simultaneously, when the target is observed in one order the background is observed in the other. The on-sky geometry shown at the bottom reveals a spatial separation of 158" between the SL1 and SL2 observations of the background and a 3".6$\times$38" (2$\times$21 pixels) overlapping aperture. NB the observations considered in this work all make use of the depicted nodding strategy.}
  \label{fig:slit}
\end{figure*}

\begin{figure*}
  \centering
  \includegraphics[width=\linewidth]{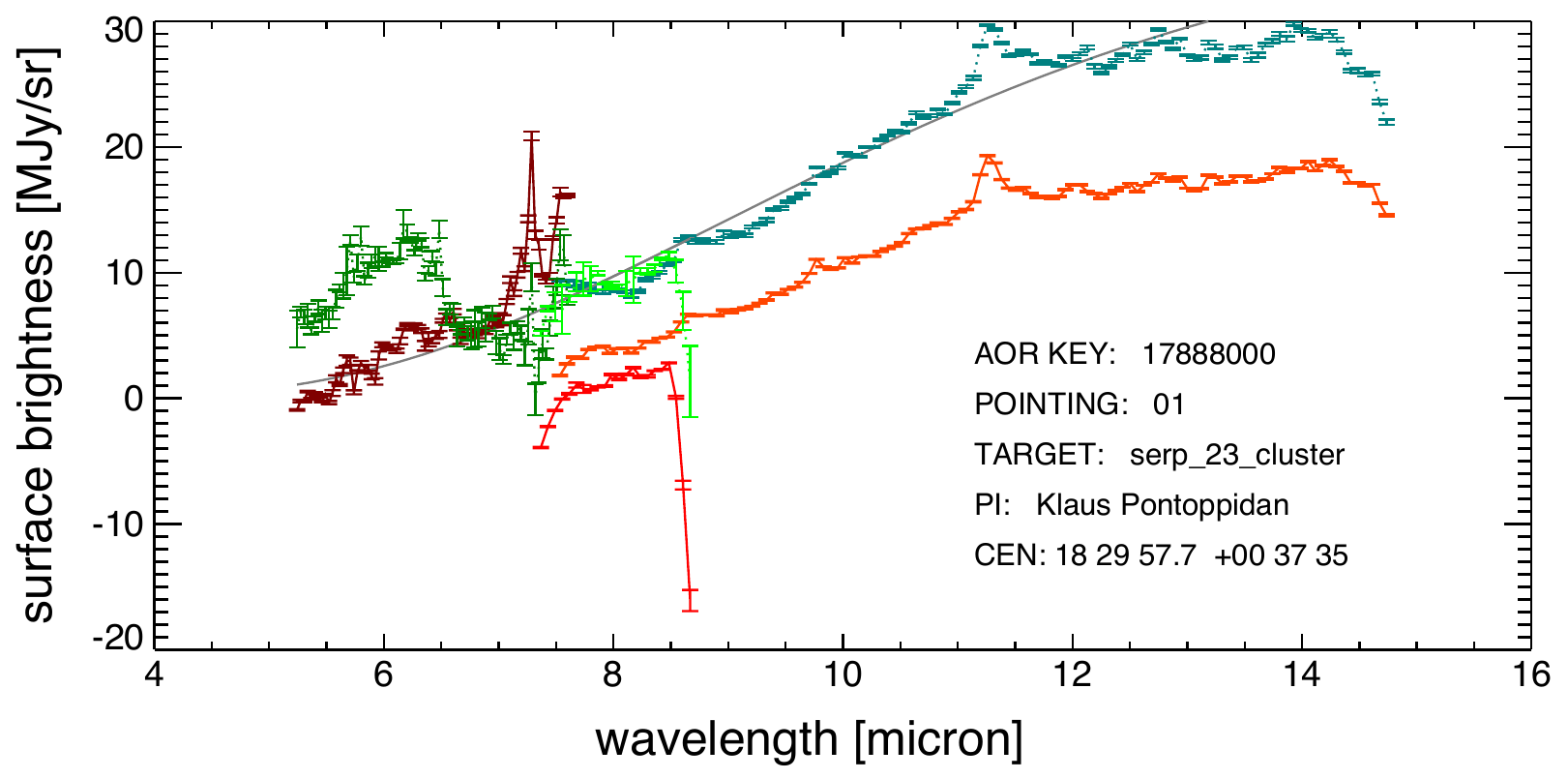}
  \caption{Extracted spectra. The SL1, 2 and 3 orders for the dark corrected spectrum have been colored light red, red and dark red, respectively. Those for the non-dark corrected spectrum light green, green and dark green, respectively. The modeled zodiacal light is shown as the grey line. See Sect.~\ref{subsec:reduction} for details.}
  \label{fig:extraction}
\end{figure*}

Next, spectra were extracted from the BCDs using the The CUbe Builder for IRS Spectra Maps (CUBISM\footnote{\url{irsa.ipac.caltech.edu/data/SPITZER/docs/dataanalysistools/tools/cubism/}}; \citealt{2007PASP..119.1133S}) tool. CUBISM was modified to allow automation of the process. Spectra were extracted using a 24x2 pixel window (see Fig.~\ref{fig:slit} for details) in the off-order position to obtain a background spectrum and saved to disk using the IPAC table format\footnote{\url{irsa.ipac.caltech.edu/applications/DDGEN/Doc/ipac\_tbl.html}} for both the dark and non-dark corrected BCDs. The associated headers track relevant information, e.g., remapped center position, target name, program name, pointing, reduction history, etc. The IPAC-formatted tables for the non-dark corrected spectra have been made publicly available (see Sect.~\ref{sec:catalogue}). Figure~\ref{fig:extraction} showcases and compares typical dark and non-dark extracted spectra. Also shown is an estimate for the zodiacal light spectrum as determined by the Zodiacal Light Model available through IPAC\footnote{\url{irsa.ipac.caltech.edu/data/SPITZER/docs/dataanalysistools/tools/contributed/general/zodiacallight/}}. The figure shows that most of the non-dark emission can be attributed to zodiacal light.

\begin{figure*}
  \centering
  \includegraphics[width=0.5\linewidth]{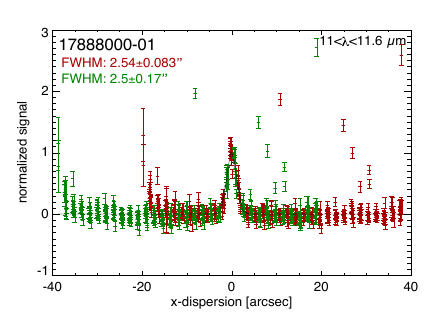}\hfill\includegraphics[width=0.5\linewidth]{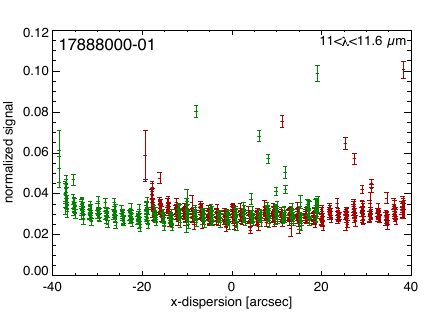}
  \caption{Determined and fitted cross-dispersion profiles for the 11.2~{\textmu}m PAH band. The two colors represent the two different nod positions. Left: Science target. Right: Background.}
  \label{fig:dispersion}
\end{figure*}

Lastly, on- and off-order cross-dispersion profiles are constructed for emission between 11-11.6~{\textmu}m. First, for each target and pointing the Data Collection Event (DCE) files for a single nod position are combined for the non-dark subtracted BCDs, FUNCs, and BMASKs. Second, the resulting combined images are collapsed in the cross-dispersion direction after straightening the source trace to achieve a sub-pixel sampling of the profile. Third, wavelength and order are associated with each data point using available calibration files (\texttt{irs\_sl\_wavesamp-[omask,wave]\_v5.fits}). Finally, the cross-dispersion profile between 11-11.6~{\textmu}m is constructed for each exposure and fitted with a Gaussian profile plus an offset. The IDL \texttt{MPFITPEAK}-procedure by Craig Markwardt is used for the fitting, forcing a positive peak and a minimum FWHM of 2".355, the extent of a point-source. Prior to fitting, the data are sigma clipped over 4 surrounding elements. Figure~\ref{fig:dispersion} provides an example of the extracted and fitted cross-dispersion profile for the same target and pointing as shown in Fig.~\ref{fig:extraction} after shifting its center to 0" and subtracting the offset. The figure reveals slightly extended PAH emission originating from the target with a FWHM of about 2".5, while the signal from the background is spatially unresolved, thus fully extended. \emph{It is noted that for all background cross-dispersion profiles the PAH signal is spatially unresolved.}

\section{Analysis}
\label{sec:analysis}

\begin{figure*}
  \includegraphics[width=\linewidth]{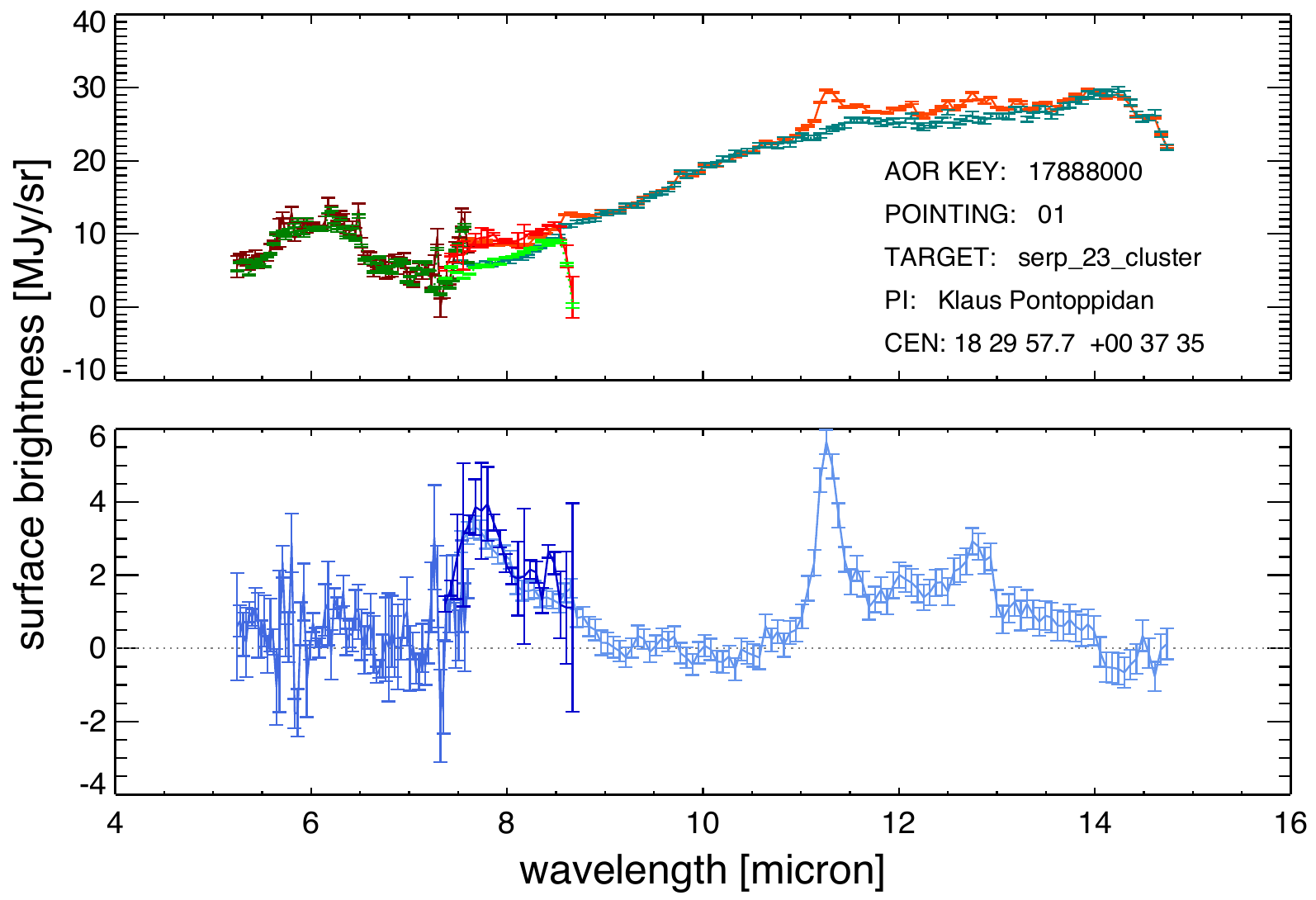}
  \caption{Top: Constructed broad band continuum (greens) matched to the non-dark corrected spectrum (reds) from Fig.~\ref{fig:extinction}. Bottom: Resulting background spectrum from subtracting the broad band continuum from the non-dark corrected spectrum. See Sect.~\ref{sec:analysis} for details.}
  \label{fig:zodiacal}
\end{figure*}

To isolate any PAH emission bands in the background spectra, a broad continuum is subtracted. This continuum primarily consists of low surface brightness zodiacal light (cf. Fig.~\ref{fig:extraction}), but at SL2 wavelengths is dominated by the IRS' detector response and appears as a broad discrete feature centered around 6~{\textmu}m. Removing this continuum is achieved in three steps. First an average zodiacal light spectrum is constructed from the non-dark subtracted spectra that fall below b = -60$^{\circ}$ using a weighted mean. The resulting zodiacal light spectrum is the average of 108 and 15 spectra with 60- and 240~s ramp times, respectively. Second, this spectrum is scaled and added to a 1$^{\rm st}$-order polynomial to match the emission at parts of the SL1 spectrum not affected by PAH emission. For the SL2 spectrum the average zodiacal spectrum is only scaled. Wavelength elements between 9-11 and beyond 13.5~{\textmu}m are used to set the continuum level for SL1, while for SL2 wavelength elements between 5.7 and 5.87~{\textmu}m are selected. Third, the continuum constructed this way is subtracted. Figure~\ref{fig:zodiacal} demonstrates this approach for the spectrum in Fig.~\ref{fig:extraction}. Next, the resulting difference spectrum, e.g., that shown in the bottom panel of Fig.~\ref{fig:zodiacal}, is integrated over intervals associated with known PAH emission. For the 12.7, 11.2 and 6.2 {\textmu}m PAH bands integration ranges are set as 12.2-13.1, 10.5-11.7 and 5.9-6.4~{\textmu}m, respectively. Because the 7.7~{\textmu}m PAH emission complex is split across SL1 and SL2 it is not considered here.

\section{Discussion}
\label{sec:discussion}

\subsection{PAHs and Dust Extinction}
\label{subsec:extinction}

The determined PAH band strengths are compared to extinction measurements, where the latter are retrieved from the Galactic Dust Reddening and Extinction service at IPAC\footnote{\url{irsa.ipac.caltech.edu/applications/DUST/docs/dustProgramInterface.html}} \citep[][]{https://doi.org/10.26131/irsa537}. The returned information has a spatial resolution of five arcminutes and includes E(B-V), 100~{\textmu}m emission, and the dust temperature. Here, the values at the reference pixel (\texttt{RefPixel}) are used. As an alternative measure, the WISE 12 micron full-sky dust map\footnote{\url{faun.rc.fas.harvard.edu/ameisner/wssa/}} is also considered. The 430, 8,000x8,000-pixel `cleaned' tiles covering the entire sky at a spatial resolution of 6".5 were obtained and the WISE12~{\textmu}m flux and standard error were computed over a 3-pixel circular aperture.

\begin{figure}
  \centering
  \includegraphics[width=\linewidth]{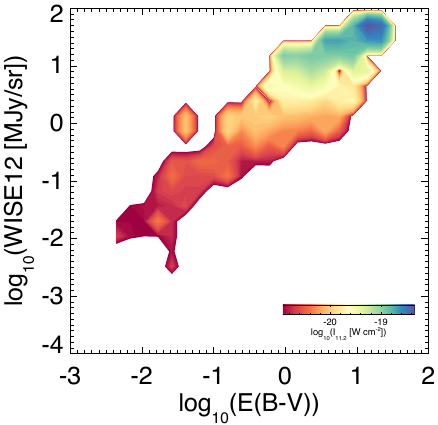}
  \caption{WISE12 band versus E(B-V) for target positions with a WISE12 band of at least 1$\times10^{\rm -4}$~MJy/sr and an associated uncertainty in the 11.2~{\textmu}m PAH band strength of less than 3$\times10^{\rm -21}$~W cm$^{\rm -2}$. The 11.2~{\textmu}m PAH band strengths are shown using (filled) contours (logarithmically scaled).}
  \label{fig:extinction}
\end{figure}

Figure~\ref{fig:extinction} shows the strength of the WISE12 band versus E(B-V) for background positions with a WISE12 band of at least 1$\times10^{\rm -4}$~MJy/sr and an associated uncertainty in the 11.2~{\textmu}m PAH band strength of less than 3$\times10^{\rm-21}$~W cm$^{\rm -2}$. The 11.2~{\textmu}m PAH band strengths are shown as the (filled) contours. The contours have been constructed from a 21x21 pixel image of the average 11.2~{\textmu}m PAH band strength of the points falling in a single pixel. The figure establishes the known correlation between WISE12 and E(B-V) and shows that the 11.2~{\textmu}m PAH band strength correlates as well.

\begin{figure*}
  \centering
  \includegraphics[width=0.5\linewidth]{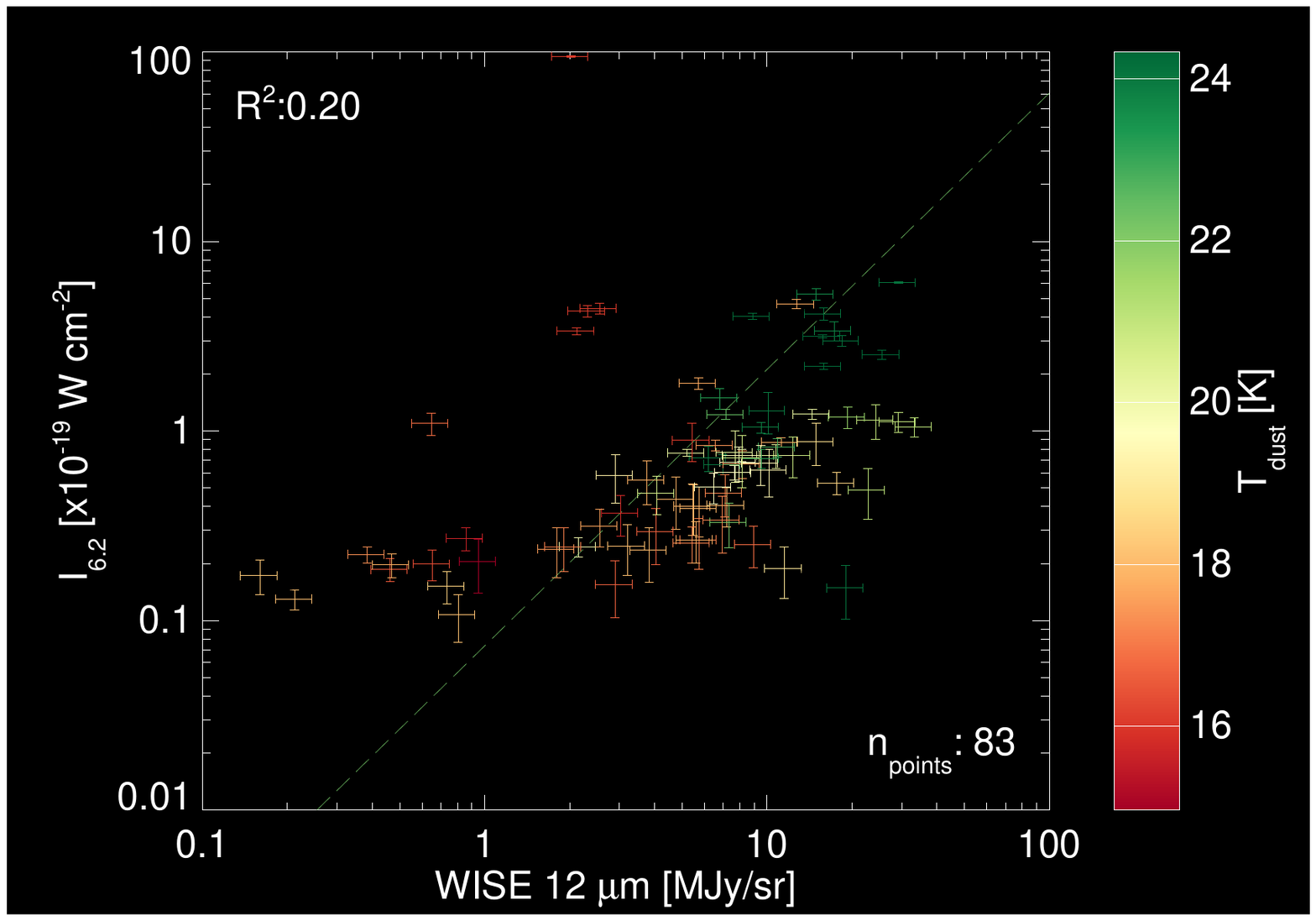}\hfill\includegraphics[width=0.5\linewidth]{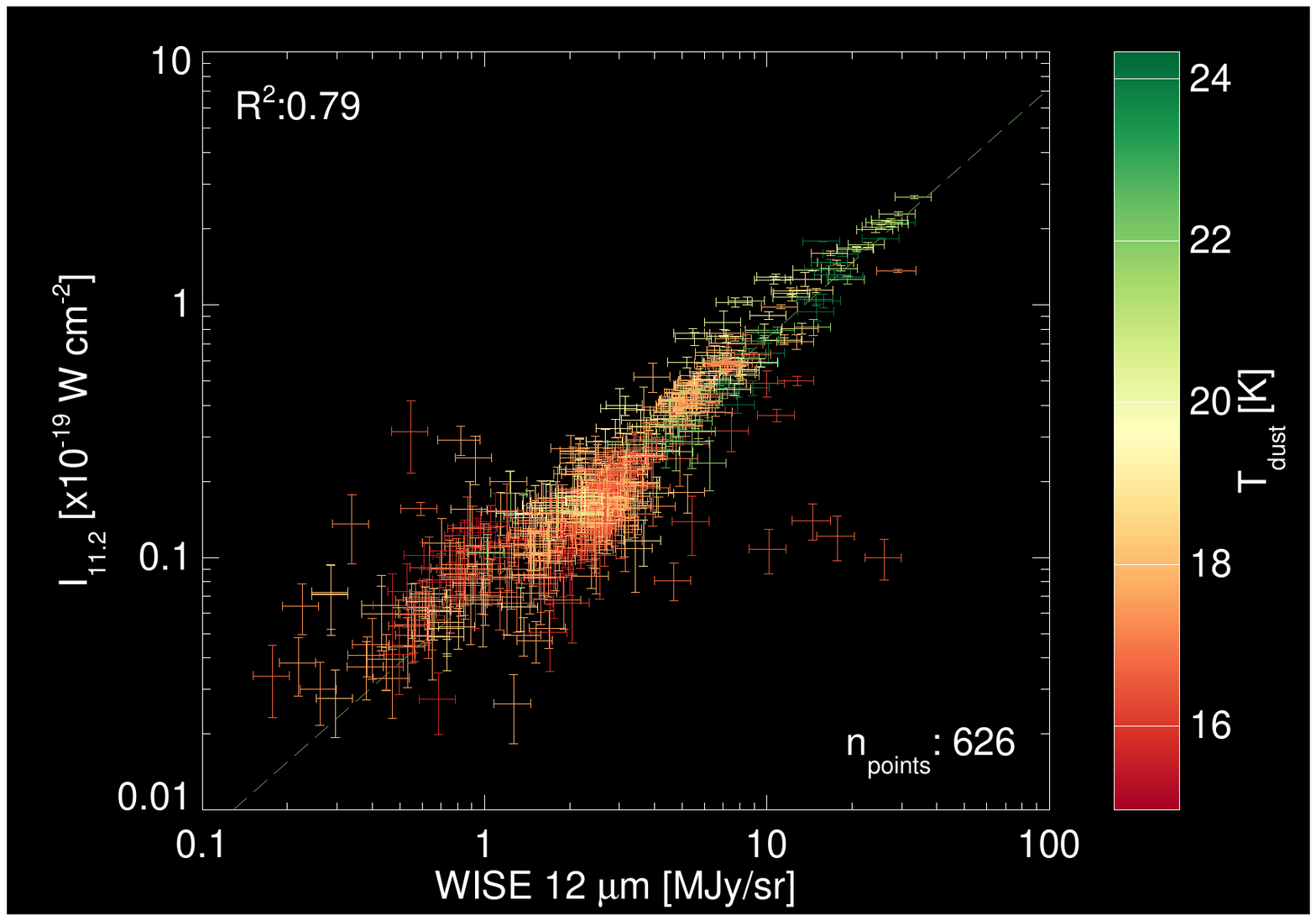}\\
  \includegraphics[width=0.5\linewidth]{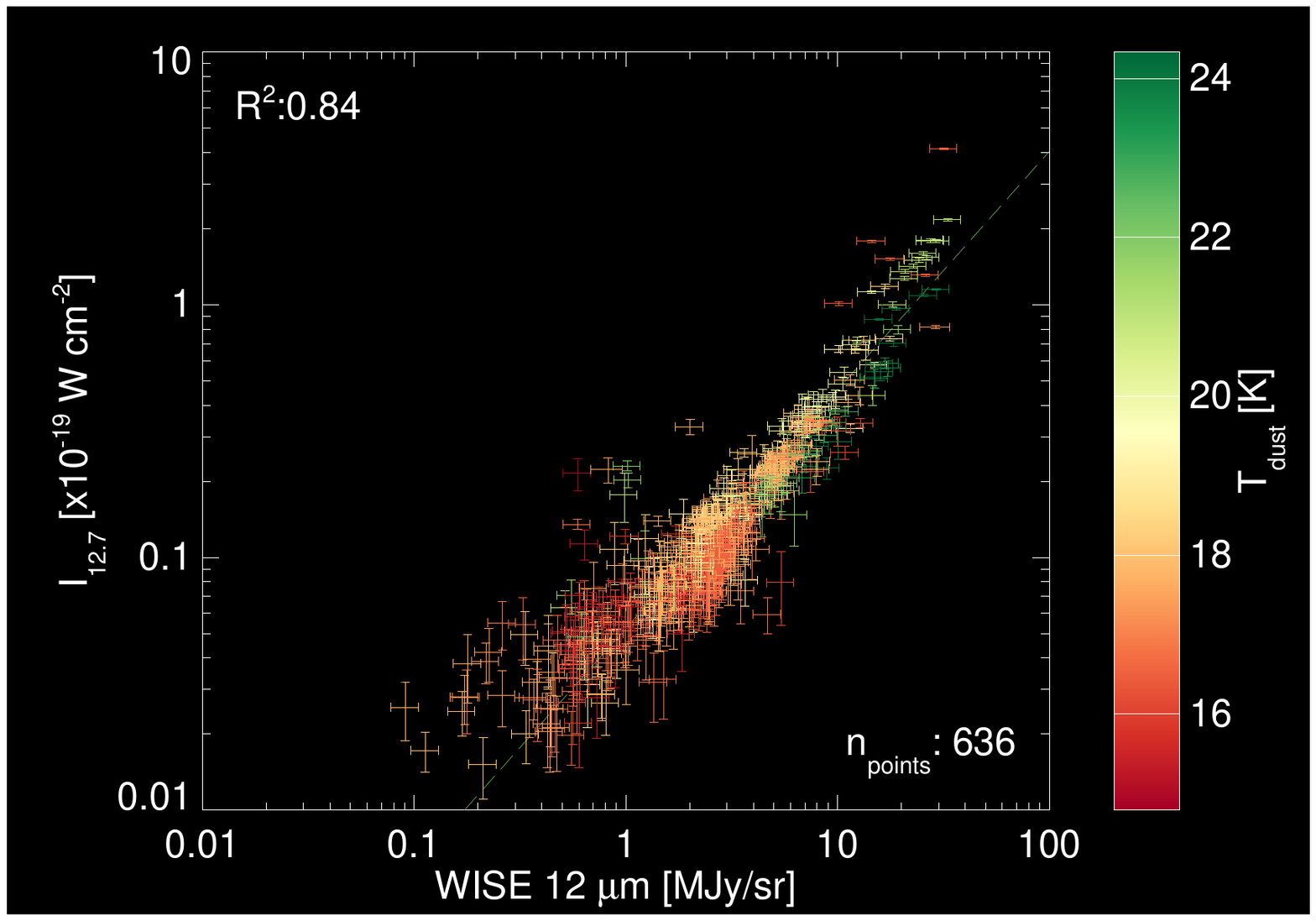}\hfill\includegraphics[width=0.5\linewidth]{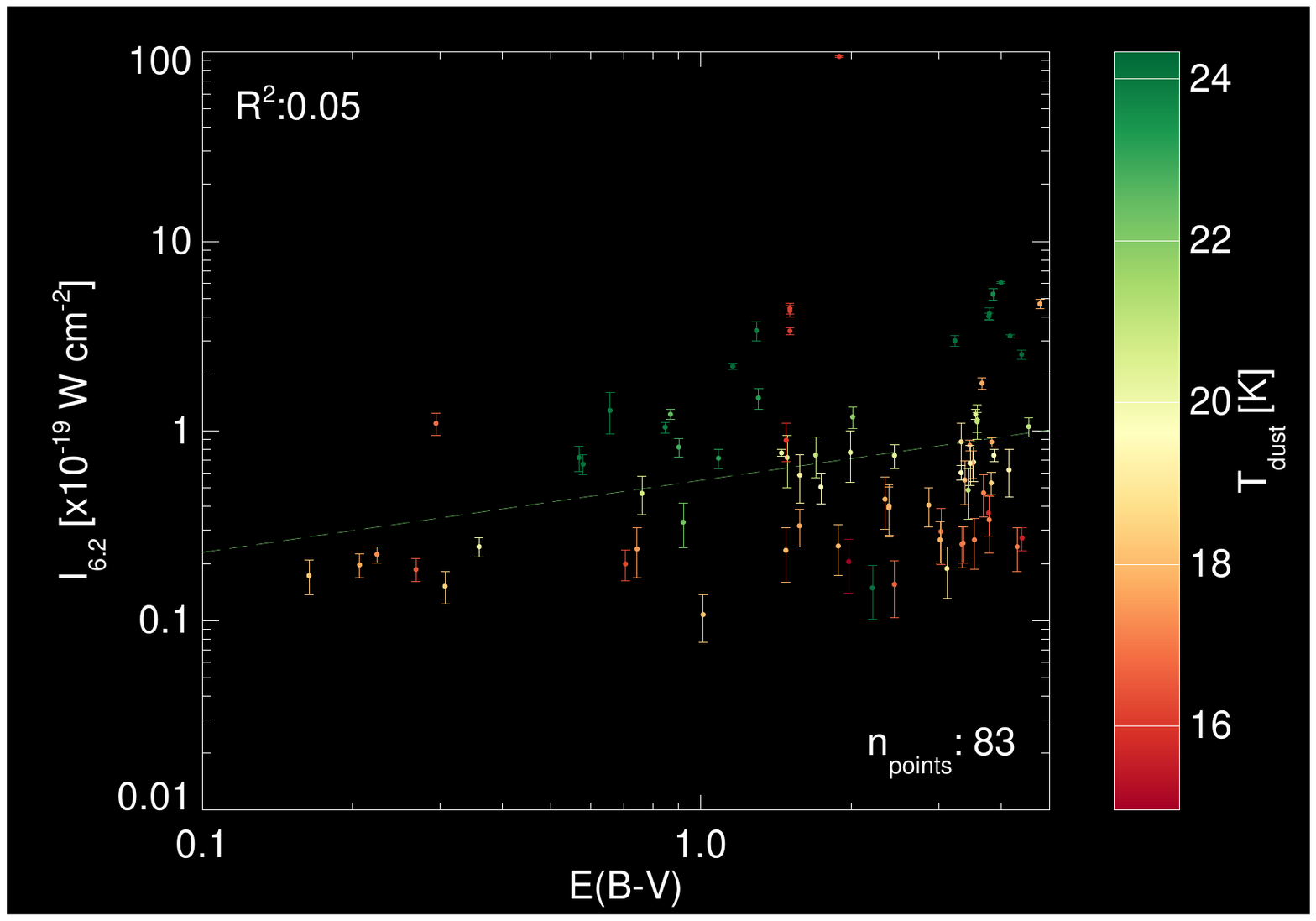}\\
  \includegraphics[width=0.5\linewidth]{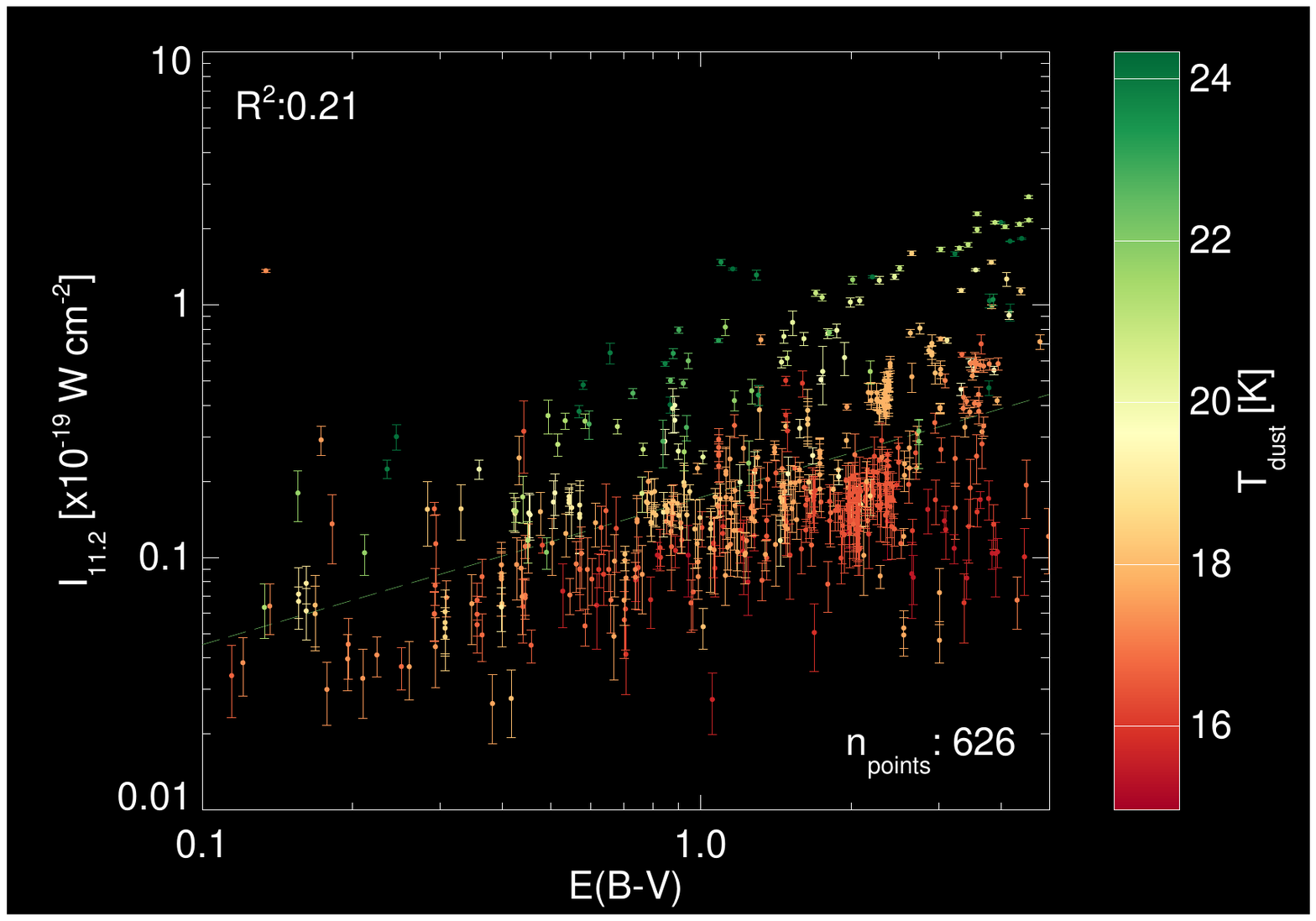}\hfill\includegraphics[width=0.5\linewidth]{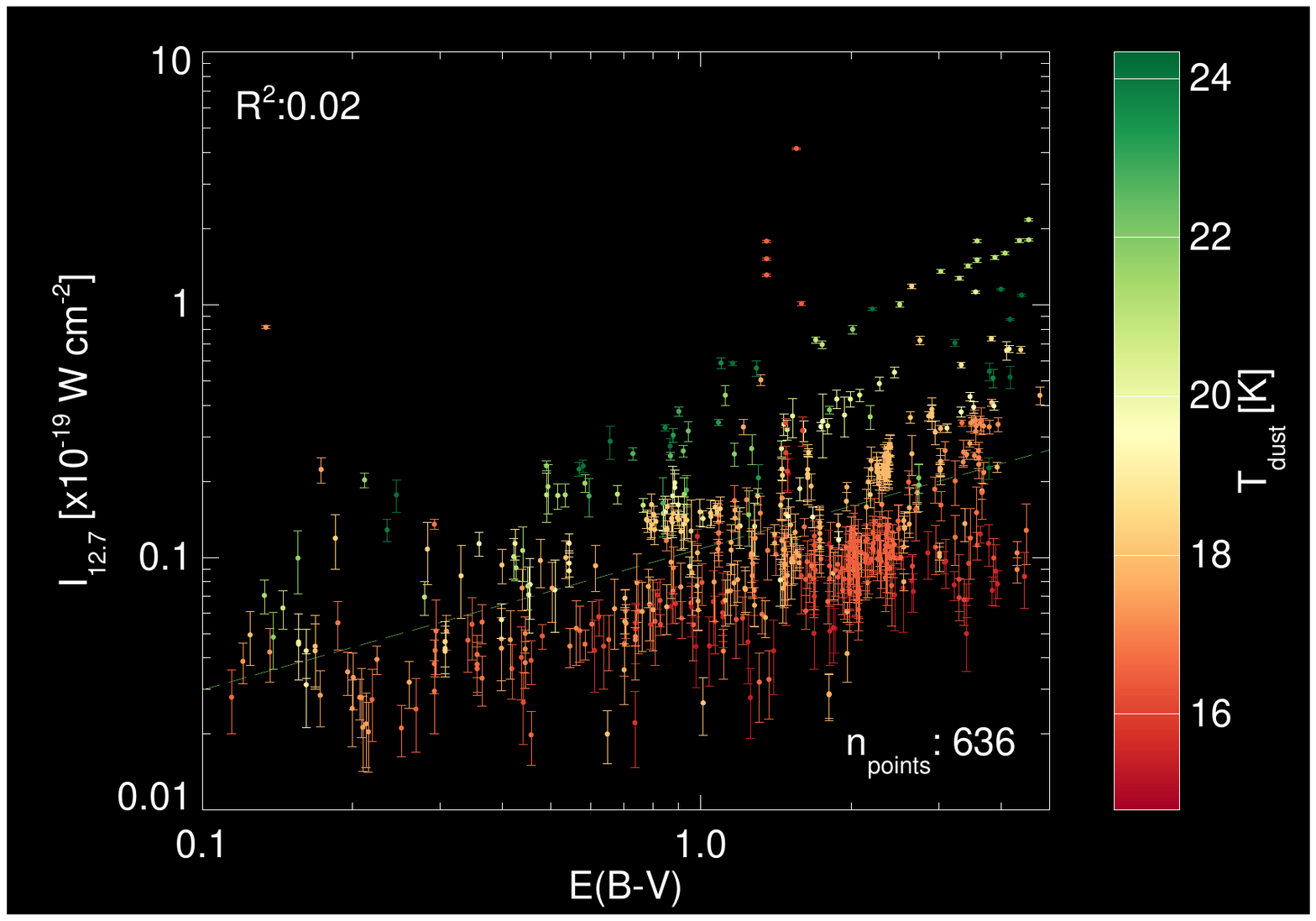}\\
  \caption{Correlation plots involving either WISE12 or E(B-V) and the 6.2, 11.2 and 12.7~{\textmu}m PAH band strengths. Each data point is color-coded according to its associated dust temperature. The data have a SNR $\geq$ 3 and 0.06 $\leq$ E(B-V) $\leq$ 5.0. Trend lines determined by fitting the data have been overlain. For each correlation the linear correlation coefficient R$^{\rm 2}$, taking uncertainties into account, has been provided. See Sect.~\ref{subsec:extinction} for details.}
  \label{fig:correlations}
\end{figure*}

Figure~\ref{fig:correlations} presents six correlation plots that involve either WISE12 or E(B-V) and the PAH band strengths for data with a signal-to-noise ratio (SNR) of at least 3 and 0.06 $\leq$ E(B-V) $\leq$ 5.0. In each panel the data points have been color-coded according their associated dust temperature. Trend lines determined by fitting a first-order polynomial, not taking into account uncertainties, have been overlain. For each correlation the linear correlation coefficient R$^{\rm 2}$, taking uncertainties into account, has been provided.

The figure shows that the 6.2~{\textmu}m PAH band is significantly hampered by poor SNR as can be inferred from Fig.~\ref{fig:zodiacal} and the top-left and middle-right panels of Fig.~\ref{fig:correlations}, where only 83 data points have a SNR$ \geq$ 3. In contrast, the correlations involving the other two PAH band intensities count almost ten times as many points. The correlations of the 11.2 and 12.7~{\textmu}m PAH band intensities with the WISE12 band are far more significant with a R$^{\rm 2}$ of 0.79 and 0.84, respectively. In addition, the correlations are quite tight and show, overall, an increase in dust temperature with an increase in both the PAH band intensity and WISE12 band.

\begin{figure*}
    \includegraphics[width=0.5\linewidth]{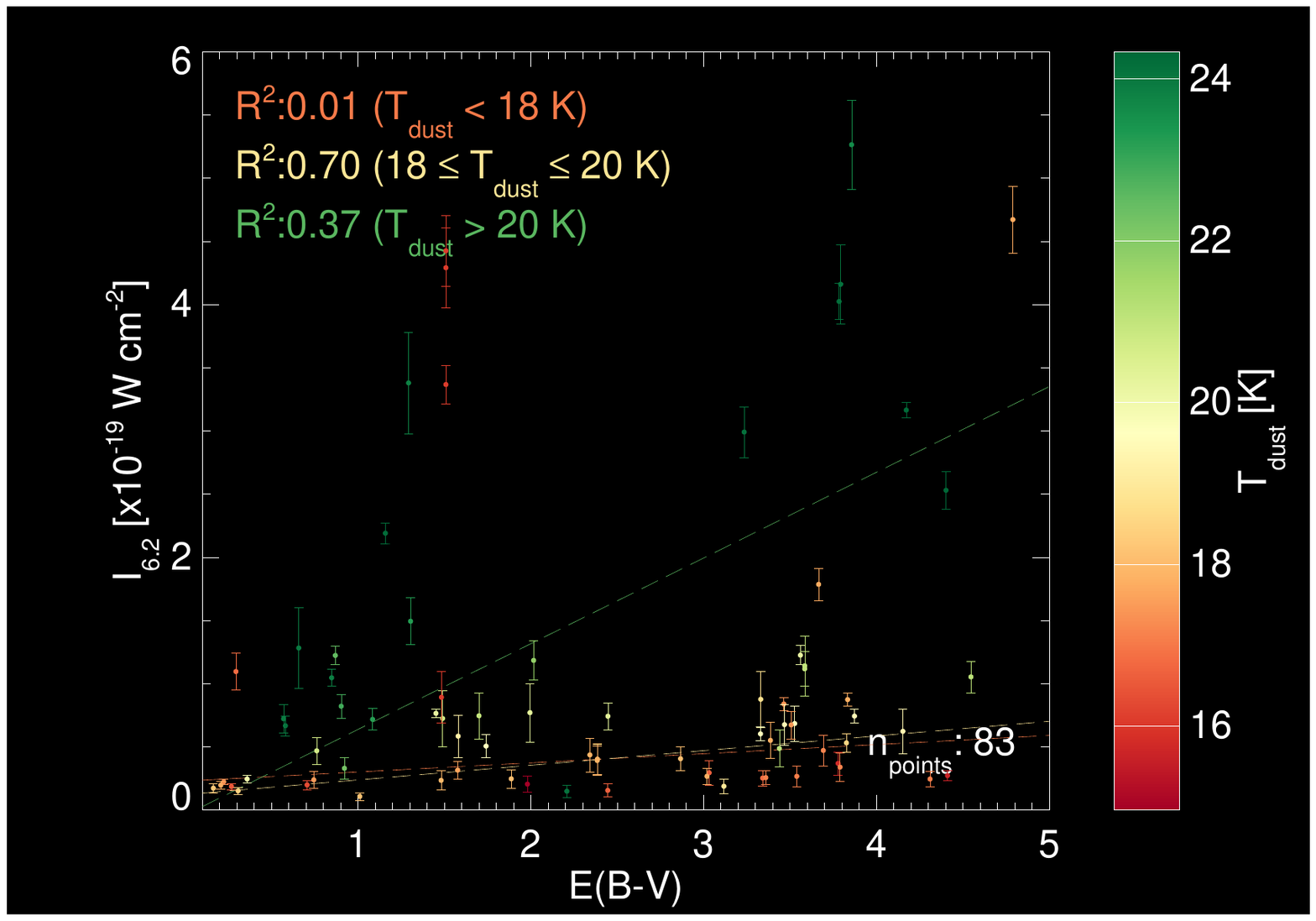}\hfill\includegraphics[width=0.5\linewidth]{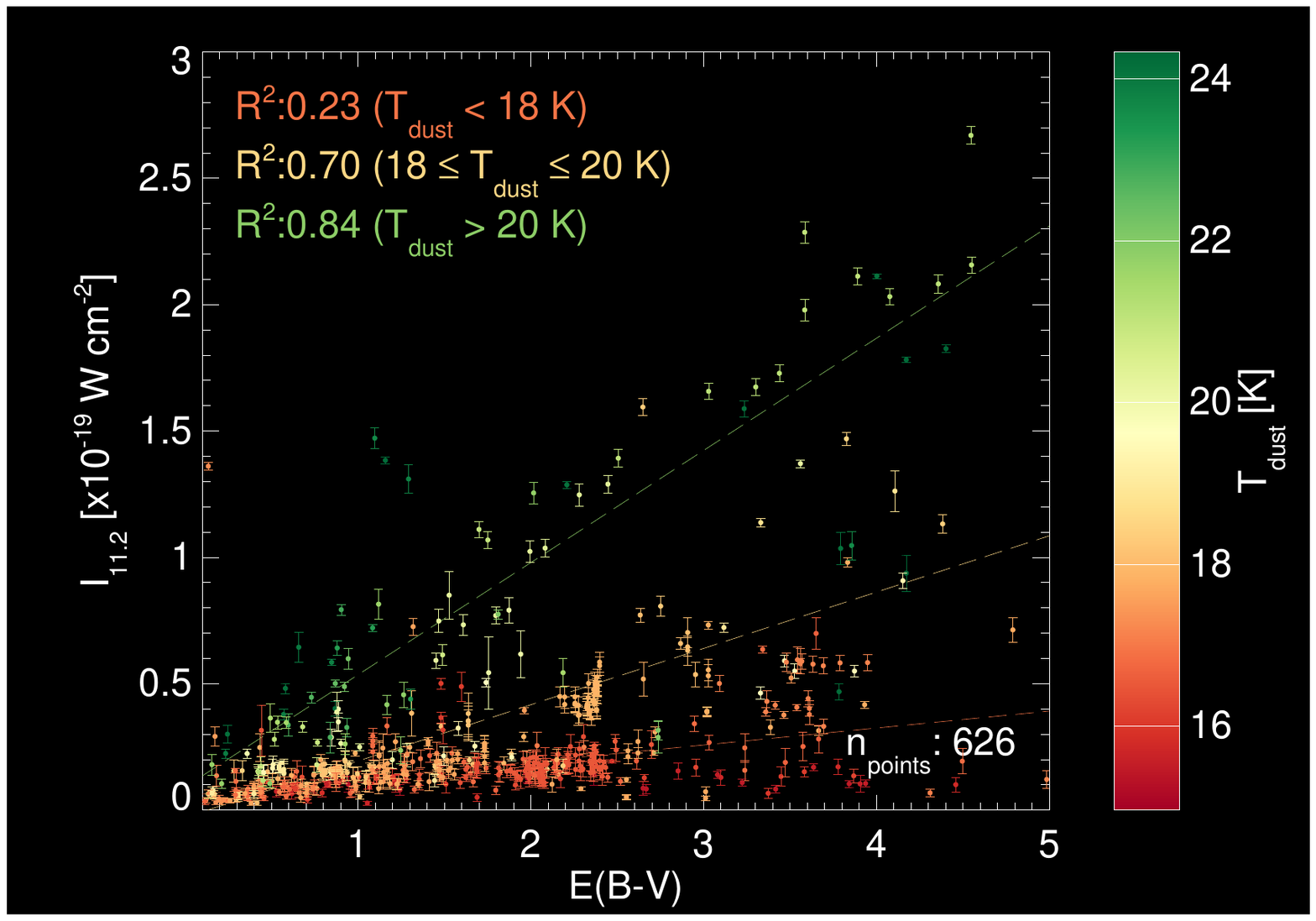}\\
    \includegraphics[width=0.5\linewidth]{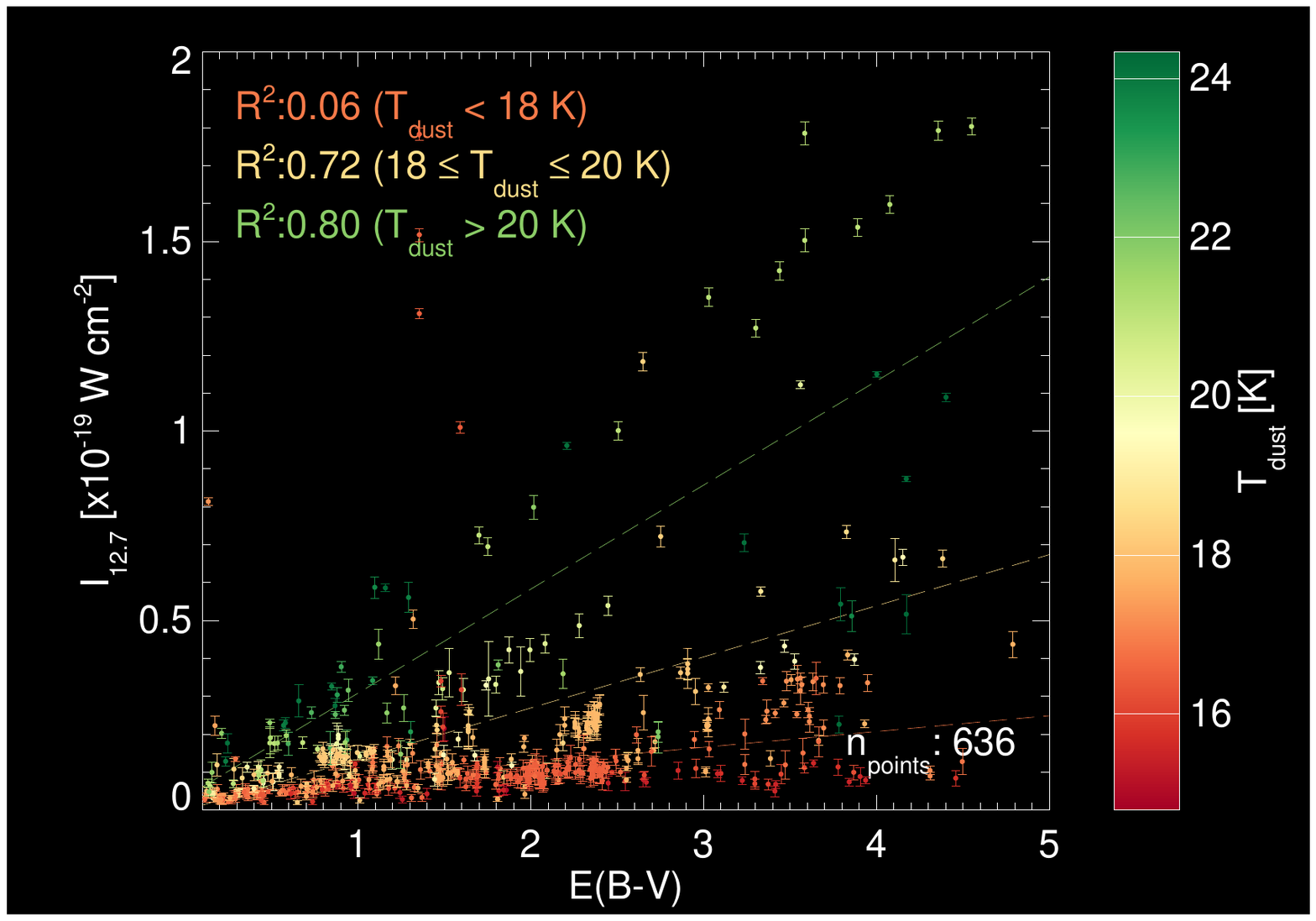}
    \caption{Correlation of the 6.2 (top-left), 11.2 (top-right), and 12.7~{\textmu}m (bottom-left) PAH band strength with E(B-V) presented on a linear scale counter to the logarithmic scale used in Fig.~\ref{fig:correlations}. Each data point is color-coded according to its dust temperature. The data have a SNR $\geq$ 3 and 0.06 $\leq$ E(B-V) $\leq$ 5.0. Trend lines determined by fitting the data, taking uncertainties into account, for three separate dust temperature intervals have been overlain. For each temperature interval the linear correlation coefficient R$^{\rm 2}$, that takes uncertainties into account, has been provided. See Sect.~\ref{subsec:extinction} for details.}
    \label{fig:linear}
\end{figure*}

Turning to the correlations of the PAH band intensities with E(B-V), while the trend lines hint at tentative correlations, none are supported by their R$^{\rm 2}$ values. Though, there appears to be a stratification with an increase in PAH band strengths with dust temperature. That is, lower temperatures are associated with lower PAH band strength intensities. This is further explored in Fig.~\ref{fig:linear}, which plots the correlations with E(B-V) using a linear scale.

 When plotting the PAH band strength correlations with E(B-V) on a linear scale (Fig.~\ref{fig:linear}), stratification with dust temperature becomes quite apparent and hints at three discrete temperature regimes. To emphasize this, trend lines have been added by separately fitting the data for T$_{\rm dust}<18$, 18 $\leq$ T$_{\rm dust}\leq20$, and T$_{\rm dust}>20$~K. For all three PAH band strengths, points falling in the lowest temperature bin have the weakest correlation per R$^{\rm 2}$ and, except for 6.2~{\textmu}m, those falling in the highest bin have the strongest. The only meaningful correlation with the 6.2~{\textmu}m PAH band strength is for 18 $\leq$ T$_{\rm dust}\leq20$. Overall, the trend line for temperatures $>$20~K visually matches best the correlation of the 11.2~{\textmu}m PAH band strength with E(B-V), which is reflected by having the best R$^{\rm 2}$ (0.84).

A likely explanation for the observed bifurcation is that those background positions with a higher dust temperature, and subsequently stronger PAH band emission, are influenced by an additional radiation source rather than the interstellar radiation field alone. This could either be because they are still receiving some portion of the radiation from the on-target source or the off-target background position happens to fall near a not too distant radiation source in, for example, a crowded star-forming region like the Orion Molecular Cloud.

\begin{figure}
  \centering
  \includegraphics[width=\linewidth]{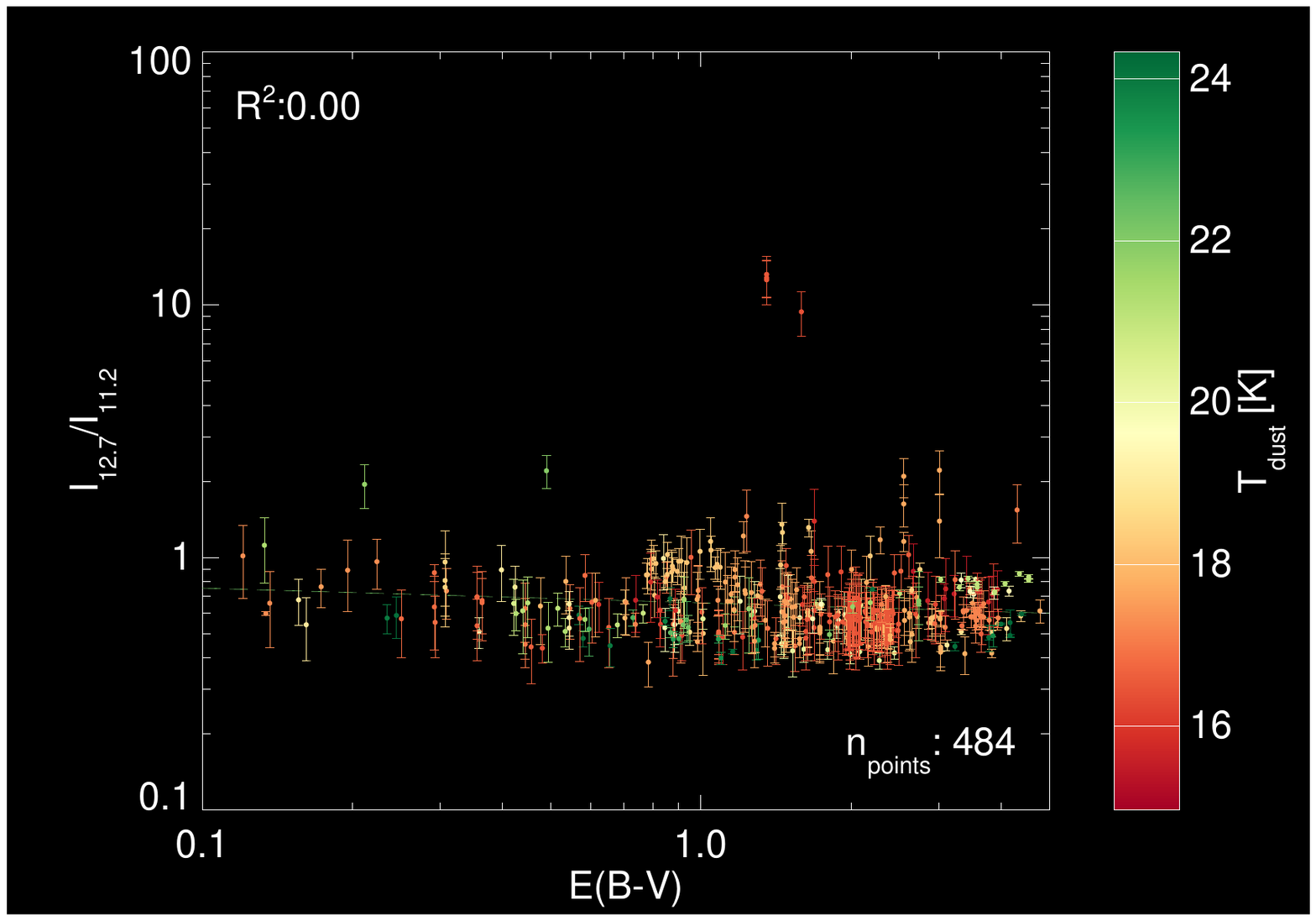}
  \caption{The 12.7/11.2~{\textmu}m PAH band strength ratio versus E(B-V). Each data point is color-coded according to its associated dust temperature. The data have a SNR $\geq$ 3 and 0.06 $\leq$ E(B V) $\leq$ 5.0. A trend line determined by fitting the data has been overlain. Provided is, taking uncertainties into account, the linear correlation coefficient R$^{\rm 2}$. See Sect.~\ref{subsec:extinction} for details.}
  \label{fig:ratio}
\end{figure}

Figure~\ref{fig:ratio} presents the 12.7/11.2~{\textmu}m PAH band strength ratio versus E(B-V). Since both bands are obtained from the same spectral segment (SL1), their ratios are relatively well constrained. Traditionally the 12.7/11.2~{\textmu}m PAH band strength ratio has been considered a measure for PAH edge structure, notably the ratio of (duo+trio)/solo-hydrogens. However, the ratio typically also shows a strong correlation with the 6.2/11.2~{\textmu}m PAH band strength ratio, which is considered a tracer for PAH charge \citep[e.g.,][]{2001A&A...370.1030H, 2014ApJ...795..110B}. The figure reveals a ratio hovering around 0.7 for all the backgrounds, with R$^{\rm 2}$ indicating no correlation, and no obvious stratification with dust temperature is discerned. This lack of stratification of the 12.7~{\textmu}m band is more fully explored in the next section.

\subsection{Emission Lines}
\label{subsec:lines}

\begin{figure}
  \centering
  \includegraphics[width=\linewidth]{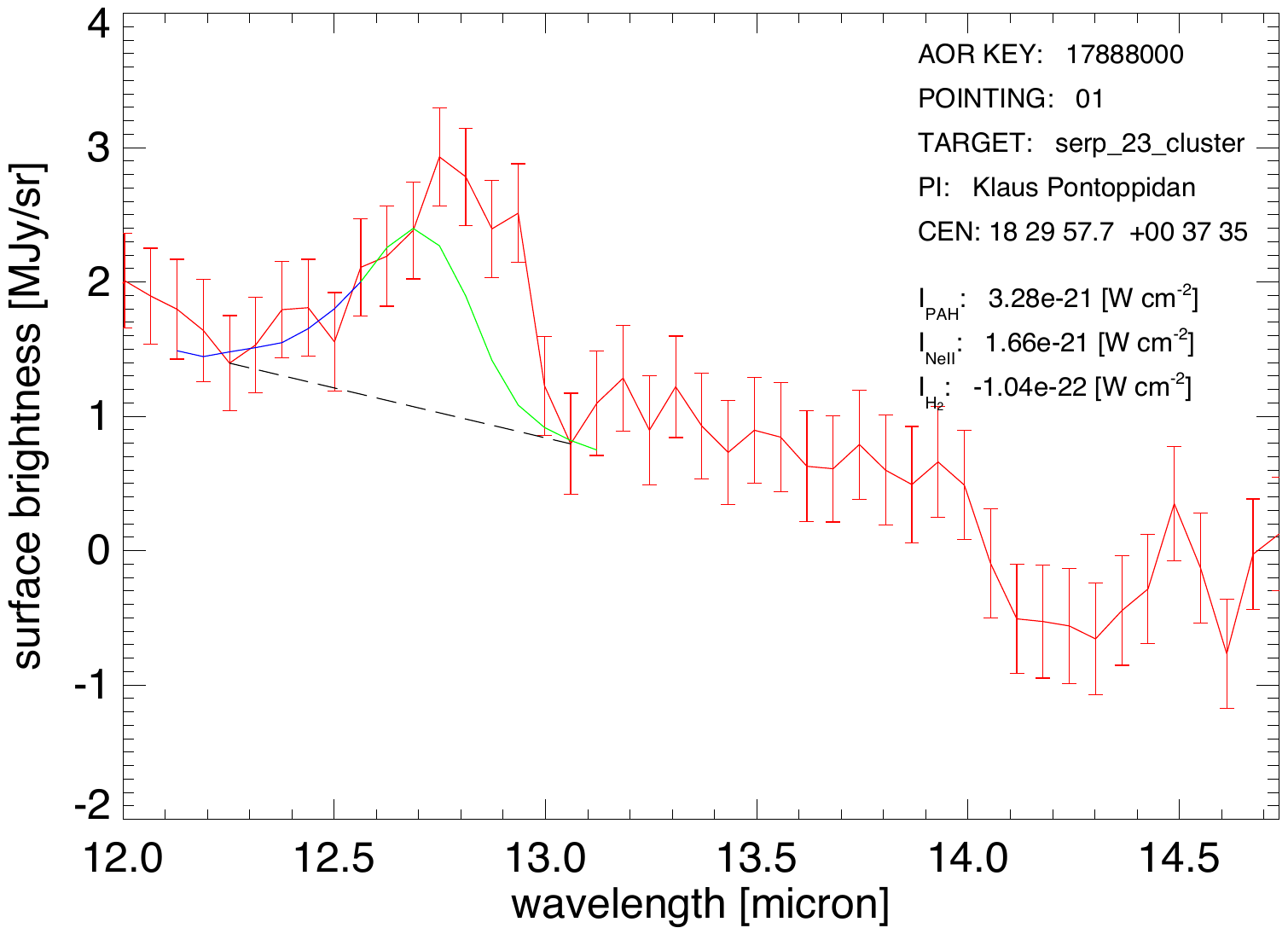}
  \caption{Disentangling Ne~II, H$_{\rm 2}$, and 12.7~{\textmu}m PAH band emission by establishing a straight-line continuum (grey-dashed line) and fitting a scaled generic 12.7~{\textmu}m PAH profile (blue- and green-solid line) from the RN NGC~7023. Shown is the fitted profile, where any excess between the observations (red) and the blue and green is attributed to H$_{\rm 2}$ S(2) and Ne~II emission, respectively. See Sect.~\ref{subsec:lines} for details.}
  \label{fig:decomposition}
\end{figure}

For many astronomical objects, the 12.7~{\textmu}m PAH emission band blends with the 12.8~{\textmu}m Ne~II and the 12.3~{\textmu}m H$_{\rm 2}$ S(2) lines. These contributions to the 12.7~{\textmu}m PAH band are removed using the approach from \citet{2015ApJ...811..153S}. Here, an emission-line-free 12.7~{\textmu}m PAH band is used as a template to fit the feature. The template used is that from \citet{2018ApJ...858...67B}, which is extracted from Spitzer-IRS observations of the RN NGC~7023. The fitting region is selected such that it excludes resolution elements affected by the two emission lines. Figure~\ref{fig:decomposition} demonstrates the approach for the spectrum shown in Figs.~\ref{fig:extraction} and \ref{fig:zodiacal}. The figure shows that some of the 12.7~{\textmu}m emission can be attributed to Ne~II, while any contribution from the H$_{\rm 2}$ S(2) line could not be reliably identified (i.e., $I_{\rm H_{2}} < 0$~W cm$^{\rm -2}$).

\begin{figure*}
  \centering
  \includegraphics[width=0.5\linewidth]{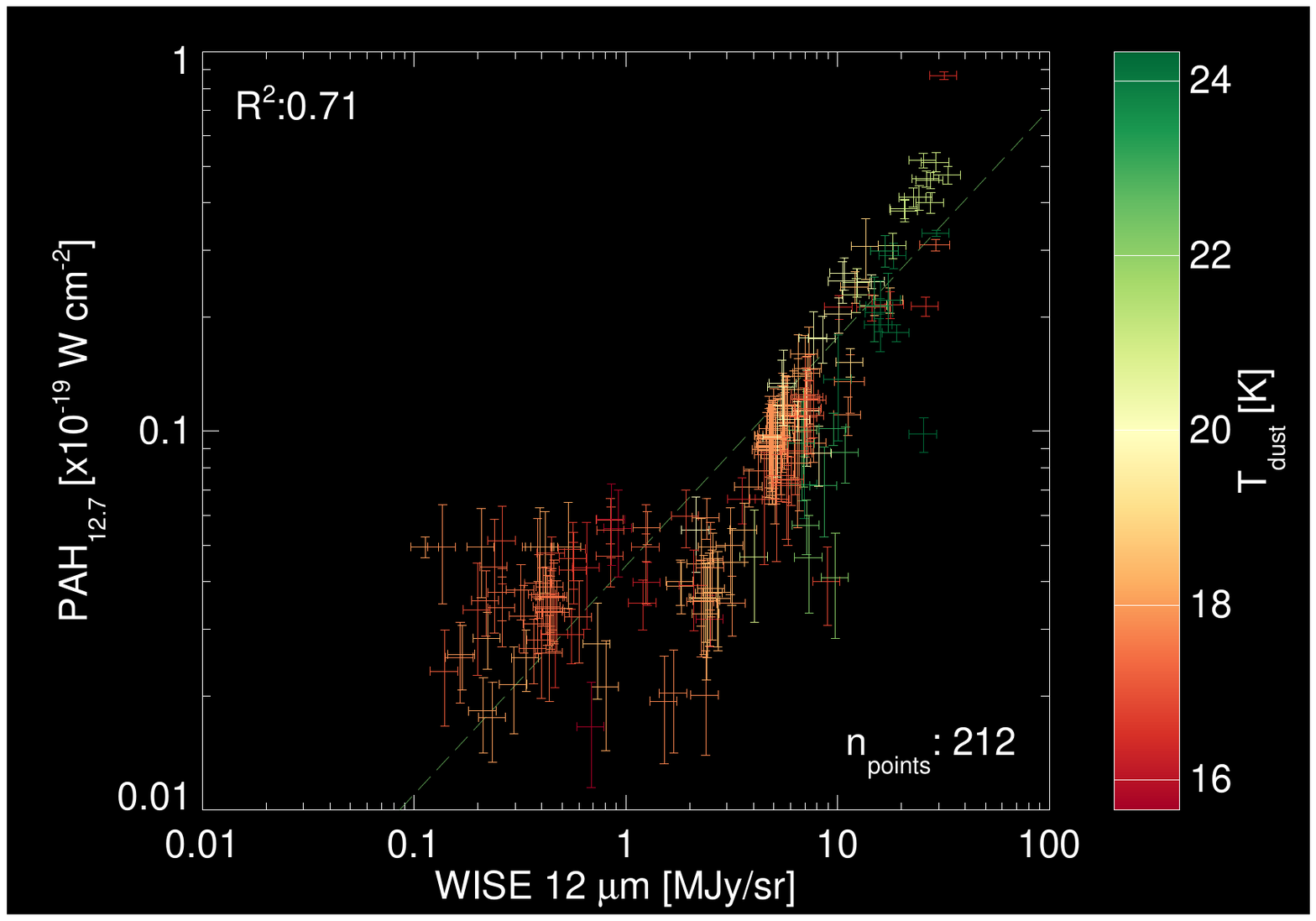}\hfill\includegraphics[width=0.5\linewidth]{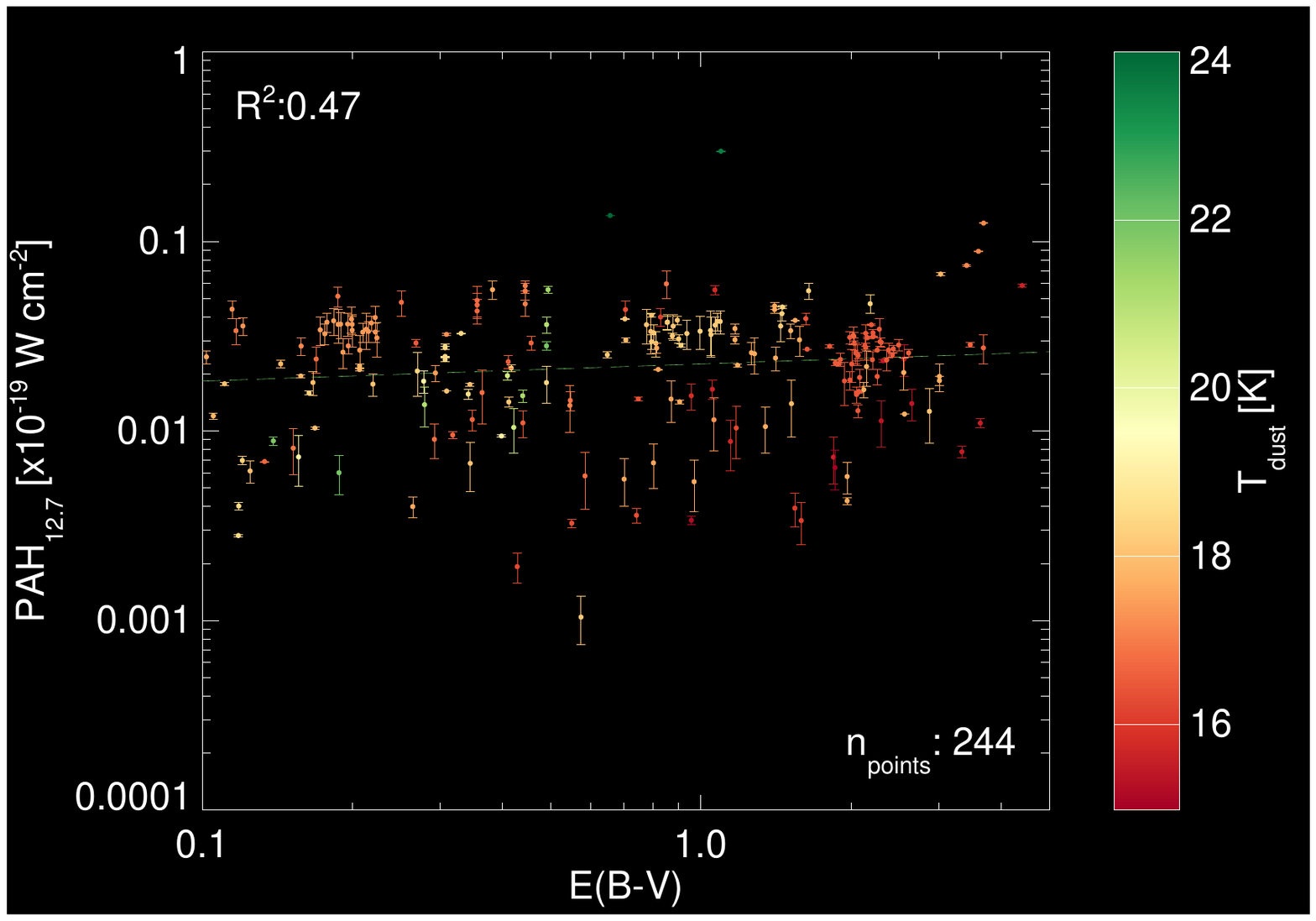}
  \caption{The correlations from Fig.~\ref{fig:correlations} involving the 12.7~{\textmu}m PAH band strength presented now using PAH$_{\rm 12.7}$ instead of I$_{\rm 12.7}$. Trend lines determined by fitting the data have been overlain. For each correlation the linear correlation coefficient R$^{\rm 2}$, that takes uncertainties into account, has been provided. See Sect.~\ref{subsec:lines} for details.}
  \label{fig:corrleations_127}
\end{figure*}

Figure~\ref{fig:corrleations_127} reproduces the correlations presented in Fig.~\ref{fig:correlations} that involve the 12.7~{\textmu}m PAH band, but now uses the strength determined from the approach described above (PAH$_{\rm 12.7}$). The figure shows that there is a drastic decrease in the number of points that pass the SNR threshold (3), with that of the PAH$_{\rm 12.7}$ versus WISE12 and E(B-V) going from 399 to 212 and 636 to 244, respectively. This would suggest that for many, if not most, background observations the emission at 12.7~{\textmu}m is affected strongly by line emission.

In terms of WISE12, above a surface brightness of $\sim$3~MJy sr$^{\rm -1}$ the correlation with PAH$_{\rm 12.7}$ seen in Fig.~\ref{fig:correlations} is maintained. However, below this value PAH$_{\rm 12.7}$ flattens out around 0.05$\times$10$^{\rm -19}$~W cm$^{\rm -2}$, albeit with a significant intrinsic spread. Though, overall the correlation remains strong with R$^{\rm 2}$ dropping only from 0.84 to 0.71. A rather unlikely explanation for the flattening out would be that the correlation seen in Fig.~\ref{fig:correlations} is initially driven solely by line emission and that any background PAH$_{\rm 12.7}$ emission is absent--Ne~II emission is typically associated with distinct astronomical objects like H~II-regions and RNe. A far more plausible explanation is that the already low SNR for these low-intensity points is not enough to confidently identify and remove any line emission. On top of that, the 12.7~{\textmu}m PAH band is not universal and the emission template taken from NGC~7023 likely not characteristic for \emph{all} background spectra.

Regarding E(B-V), the visually spurious correlation with $I_{\rm 12.7}$ present in Fig.~\ref{fig:correlations} is now entirely absent, flattened out, and hovering around 0.02$\times$10$^{\rm -19}$~W cm$^{\rm -2}$ in the PAH$_{\rm 12.7}$ case. Though, there is two orders ($\sim$0.001-0.1) of intrinsic scatter, R$^{\rm 2}$ seemingly improves from 0.02 to 0.47, and the obvious stratification with dust temperature is now gone.

\begin{figure}
  \centering
  \includegraphics[width=\linewidth]{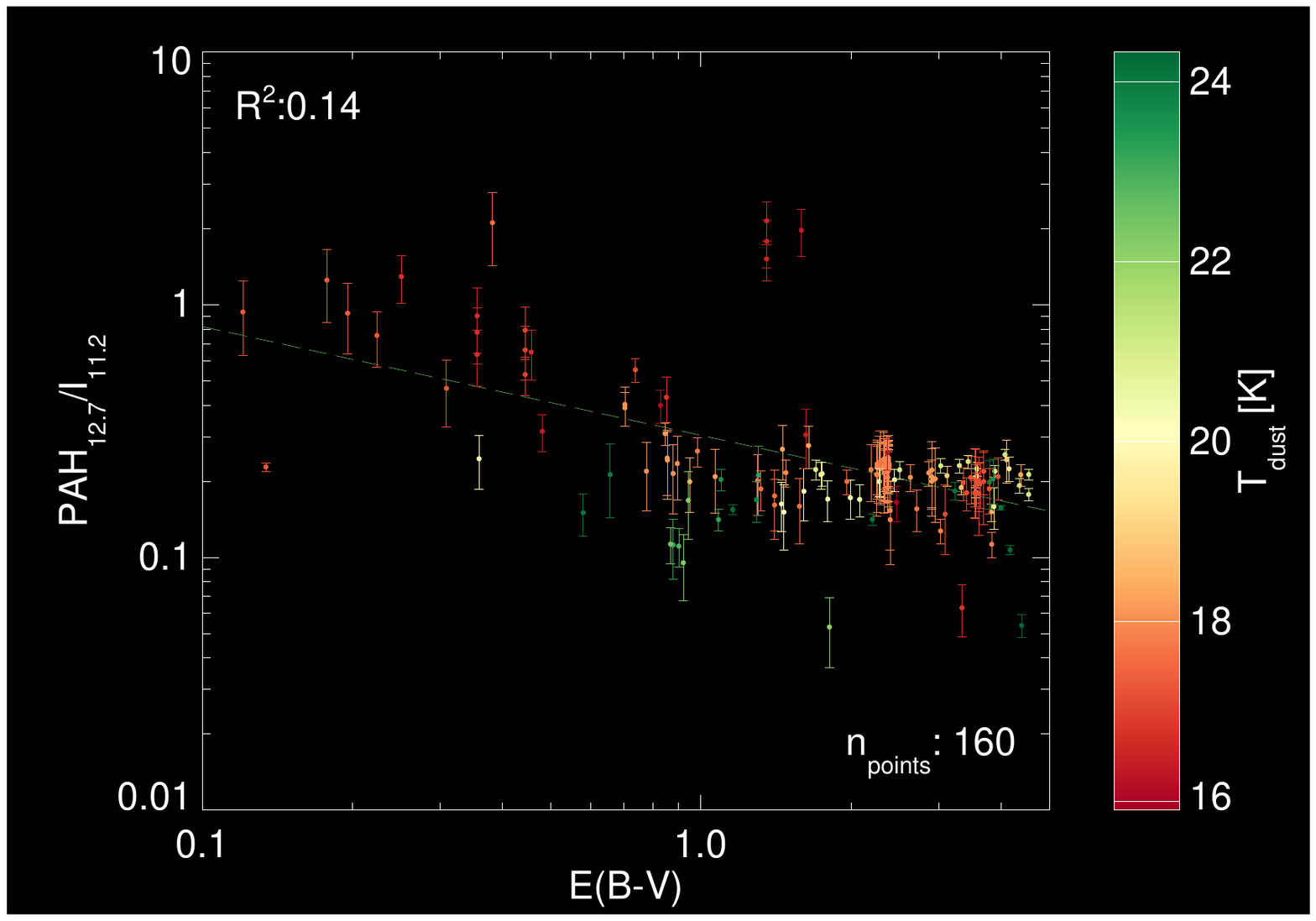}
  \caption{PAH$_{\rm 12.7}$/11.2~{\textmu}m PAH band strength ratio. Each data point is color-coded according to its associated dust temperature. The data have a SNR $\geq$ 3 and 0.06 $\leq$ E(B V) $\leq$ 5.0. A trend line determined by fitting the data has been overlain. Provided is, taking uncertainties into account, the linear correlation coefficient R$^{\rm 2}$. See Section~\ref{subsec:lines} for details.}
  \label{fig:ratio_127}
\end{figure}

Figure~\ref{fig:ratio_127} correlates E(B-V) with the PAH$_{\rm 12.7}$/11.2~{\textmu}m PAH band strength as a counterpart to Fig~\ref{fig:ratio} that uses I$_{\rm 12.7}$ instead. Again, the number of viable data points drops significantly, going from 484 down to 160. The visually apparent decrease in the ratio with E(B-V) is perhaps somewhat more prominent here, with the flattening out occurring around E(B-V) $\simeq$ 1.0 and hovering at a PAH$_{\rm 12.7}$/11.2~{\textmu}m ratio of 0.2, down from $\sim$0.7 in Fig.~\ref{fig:ratio}. Here, again R$^{\rm 2}$ seems to suggest a somewhat stronger correlation, albeit still marginal.

\subsection{Spectroscopic Database Fitting}
\label{subsec:fitting}

\begin{figure}
  \centering
  \includegraphics[width=\linewidth]{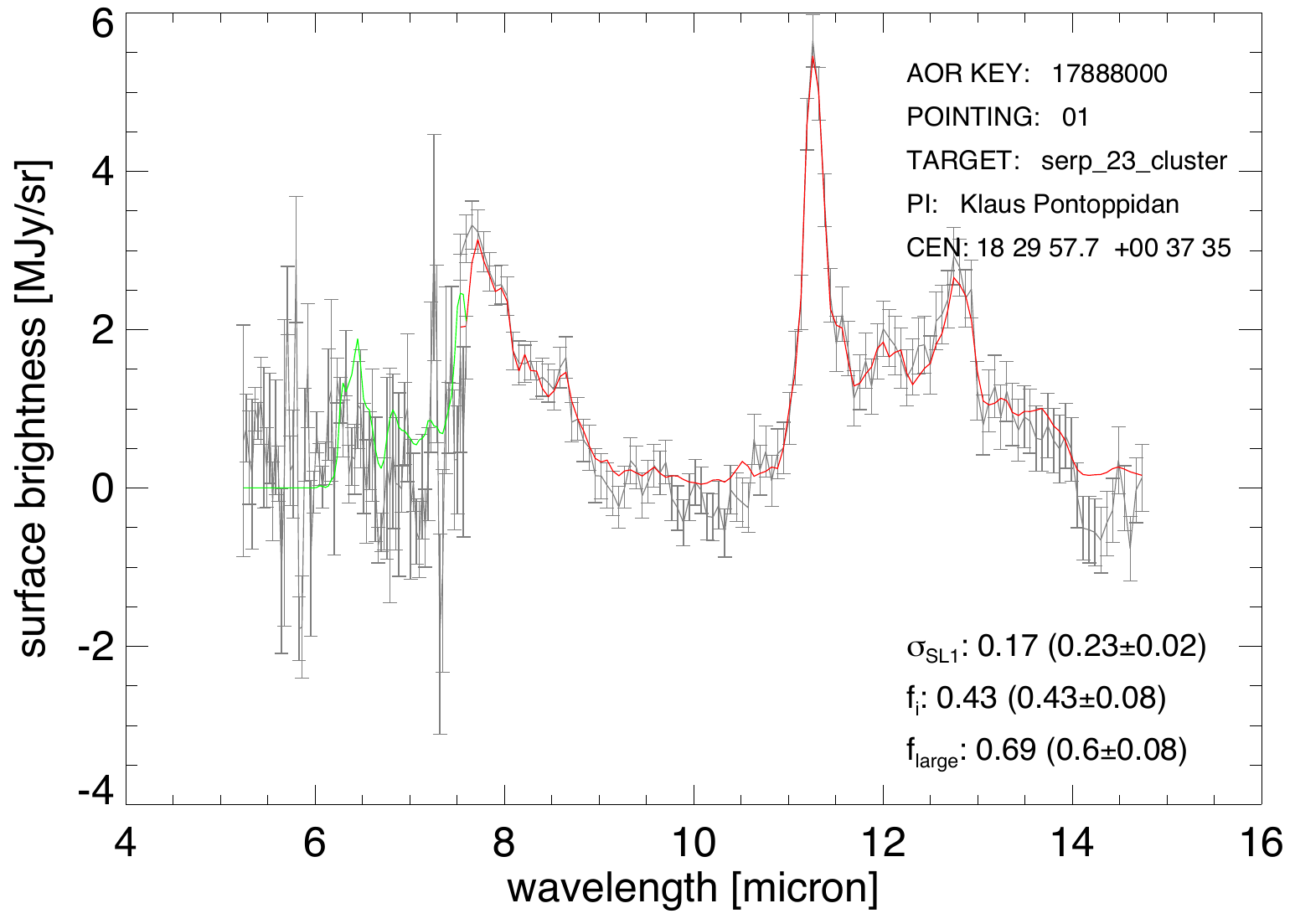}
  \caption{PAHdb-fit (Sl1=red, Sl2=green) to the background spectrum (in grey) from Fig.~\ref{fig:zodiacal}. Indicated are the single run error ($\sigma_{\rm SL1})$, cation- ($f_{\rm i}$) and large-PAH ($f_{\rm large}$) fractions and, in parenthesis, the corresponding values with their associated uncertainties from a Monte Carlo technique. See Section~\ref{subsec:fitting} for details.}
  \label{fig:pahdb_fit}
\end{figure}

Turning to the library of computed spectra at version 3.20 of the NASA Ames PAH IR Spectroscopic Database \citep[PAHdb hereafter\footnote{\url{www.astrochemistry.org/pahdb/}},][]{2010ApJS..189..341B, 2014ApJSS..211....8B, 2018ApJS..234...32B, 2020ApJS..251...22M}, PAH emission spectra are synthesized for an excitation energy of 7~eV and used to perform a fit to each background spectrum using software tools also provided by PAHdb. The propagated observational uncertainties are taken into account and, to obtain errors for the derived ionization- ($f_{\rm i}\equiv n_{\rm cation}/(n_{\rm cation}+n_{\rm neutral})$) and large-PAH fraction ($f_{\rm large}$; $N_{\rm carbon}>50$), a Monte Carlo technique is employed in which the spectra are permuted uniformly within the observational uncertainties and re-fitted 1024 times. The error ($\sigma_{\rm SL1}$) is computed as the area of the absolute value of the residual over the area of the astronomical spectra. The integration is done in frequency-space (cm$^{\rm -1}$) as it is linear in energy. See \cite{2018ApJ...858...67B} for a description of the employed modeling and a discussion of potential caveats. The mean and standard deviation of $f_{\rm i}$ and $f_{\rm large}$ are subsequently determined for each spectrum. Figure~\ref{fig:pahdb_fit} presents the results following this approach for the spectrum shown in Fig.~\ref{fig:zodiacal} and indicates both the single-run and Monte Carlo derived values for $\sigma_{\rm SL1}$, $f_{\rm i}$, and $f_{\rm large}$. Note that any potential emission lines were not removed (see Sect.~\ref{subsec:lines}). The figure shows a reasonably good fit in the SL1 region with $\sigma_{\rm SL1}$ of 0.17 for the non-perturbed case, which matches general expectations \citep[e.g.,][]{2022ApJ...931...38M}. The permuted average error is larger at 0.23$\pm$0.08 and reflects the considerable uncertainty on the data. However, $f_{\rm i}$ and $f_{\rm large}$ are more consistent at 0.43 and 0.43$\pm$0.08 and 0.69 and 0.60$\pm$0.08, respectively. Assessment of the fit to the SL2 segment of the spectrum is significantly hampered for this particular background spectrum due to its poor quality, which is unfortunately also the case for most of the other background spectra.

\begin{figure*}
  \centering
  \includegraphics[width=0.5\linewidth]{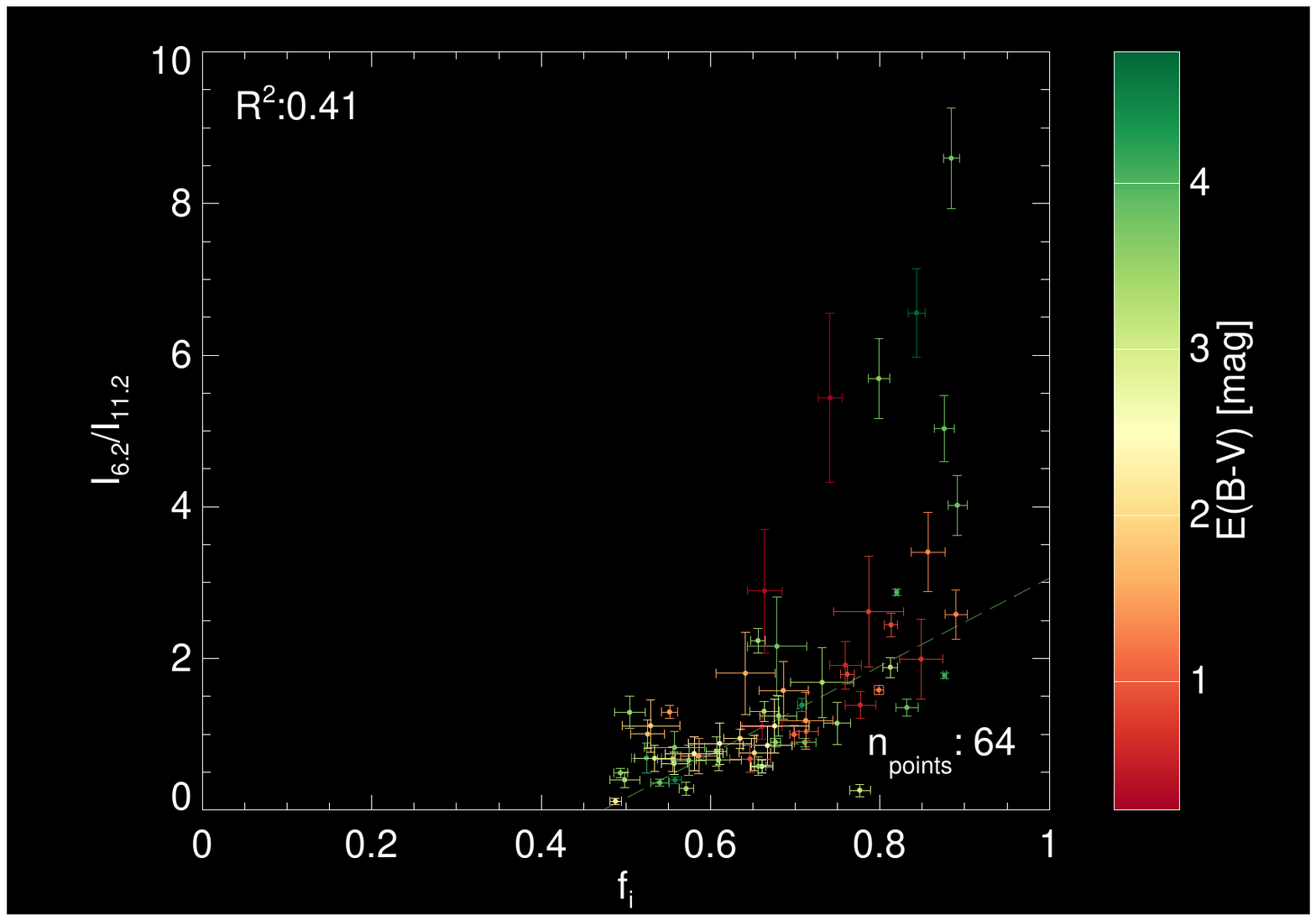}\hfill\includegraphics[width=0.5\linewidth]{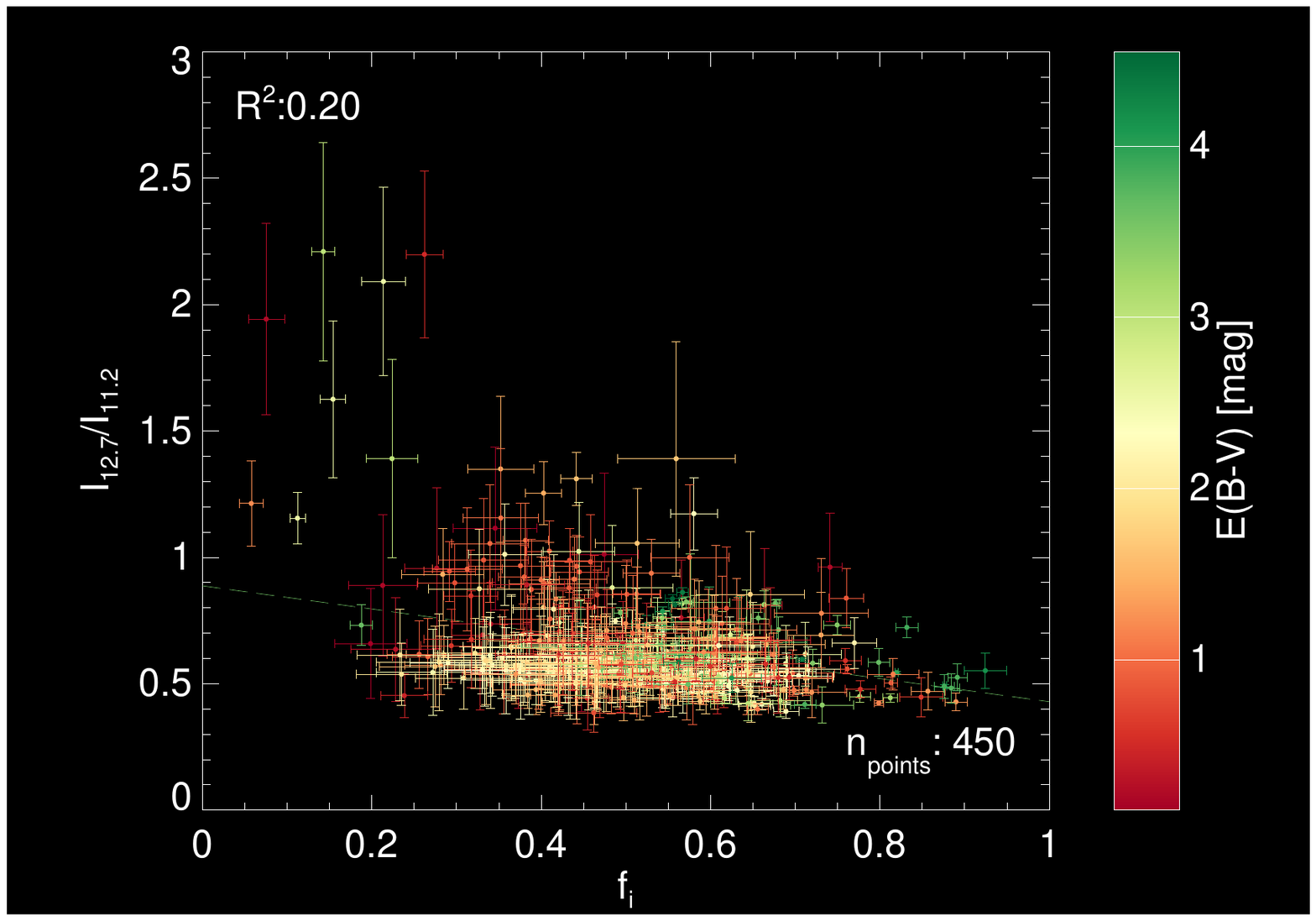}
  \includegraphics[width=0.5\linewidth]{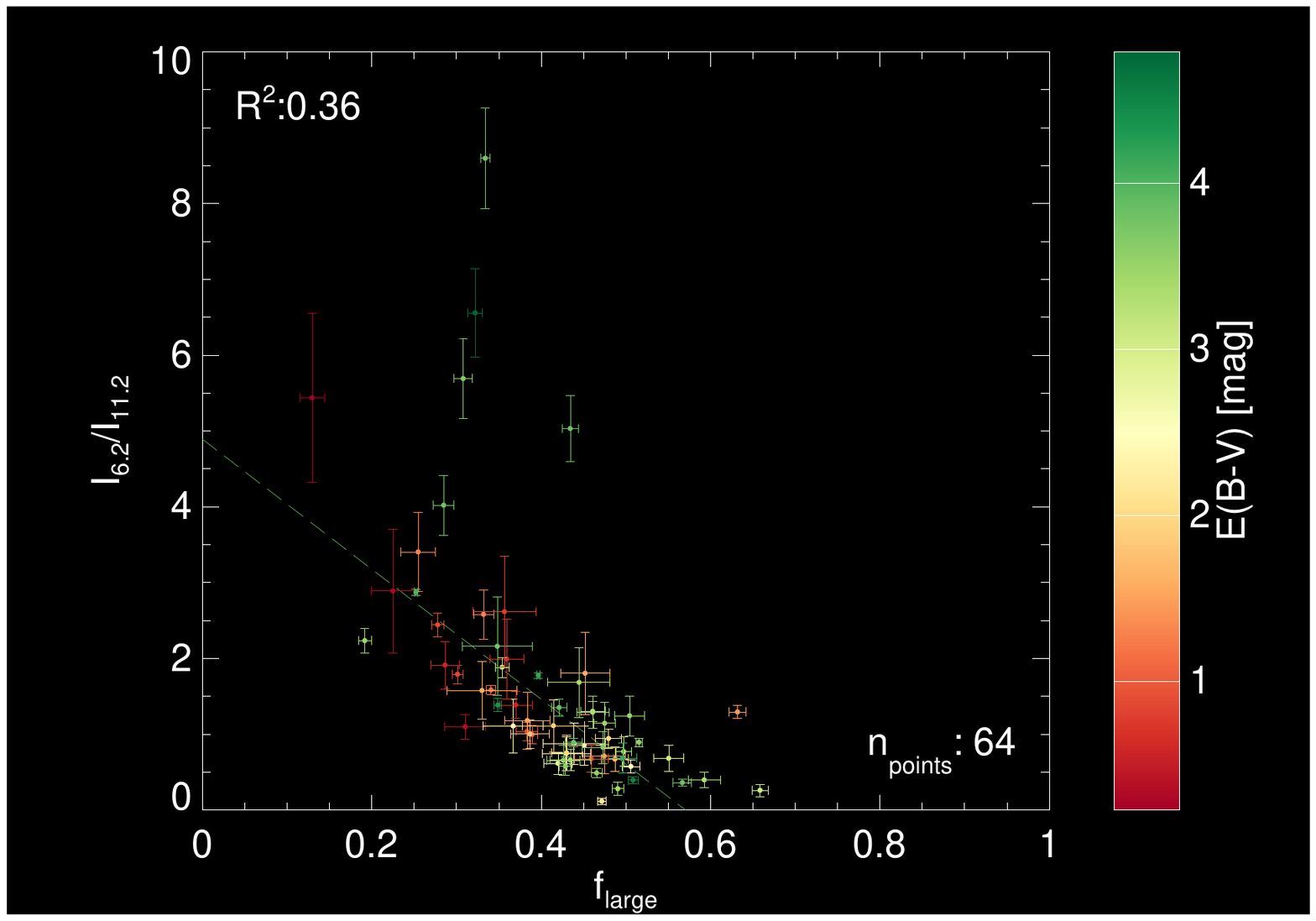}\hfill\includegraphics[width=0.5\linewidth]{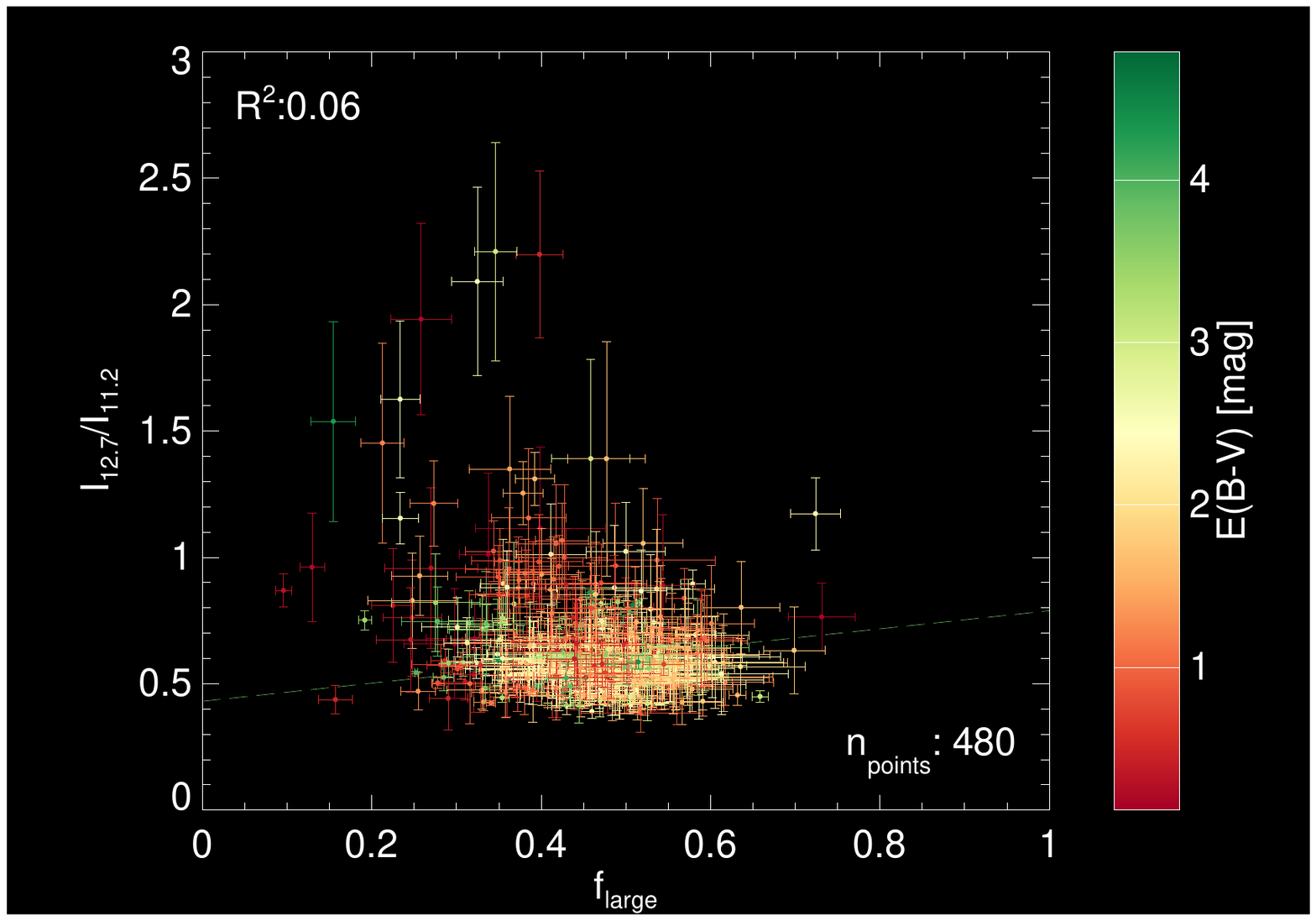}
  \includegraphics[width=0.5\linewidth]{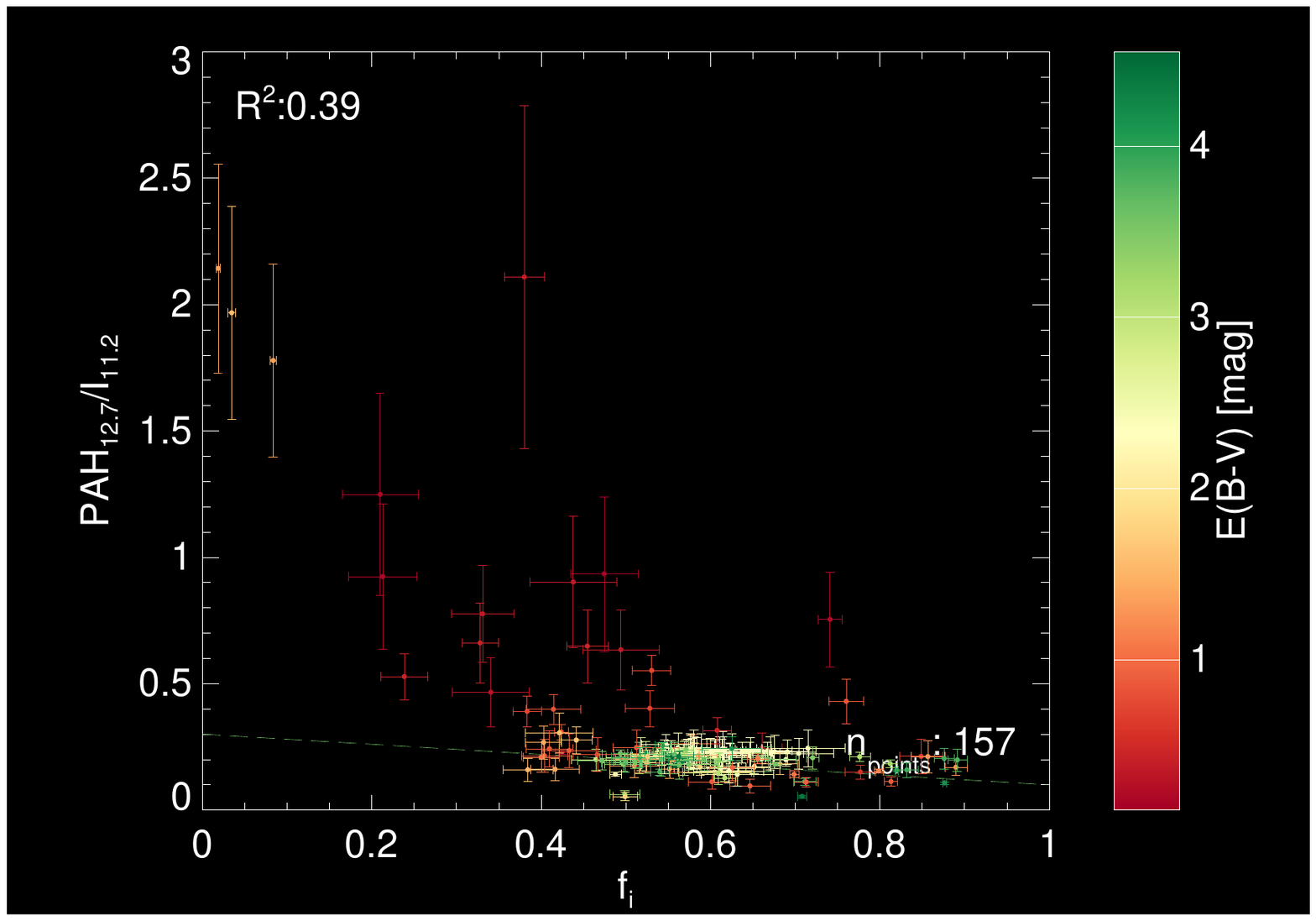}\hfill\includegraphics[width=0.5\linewidth]{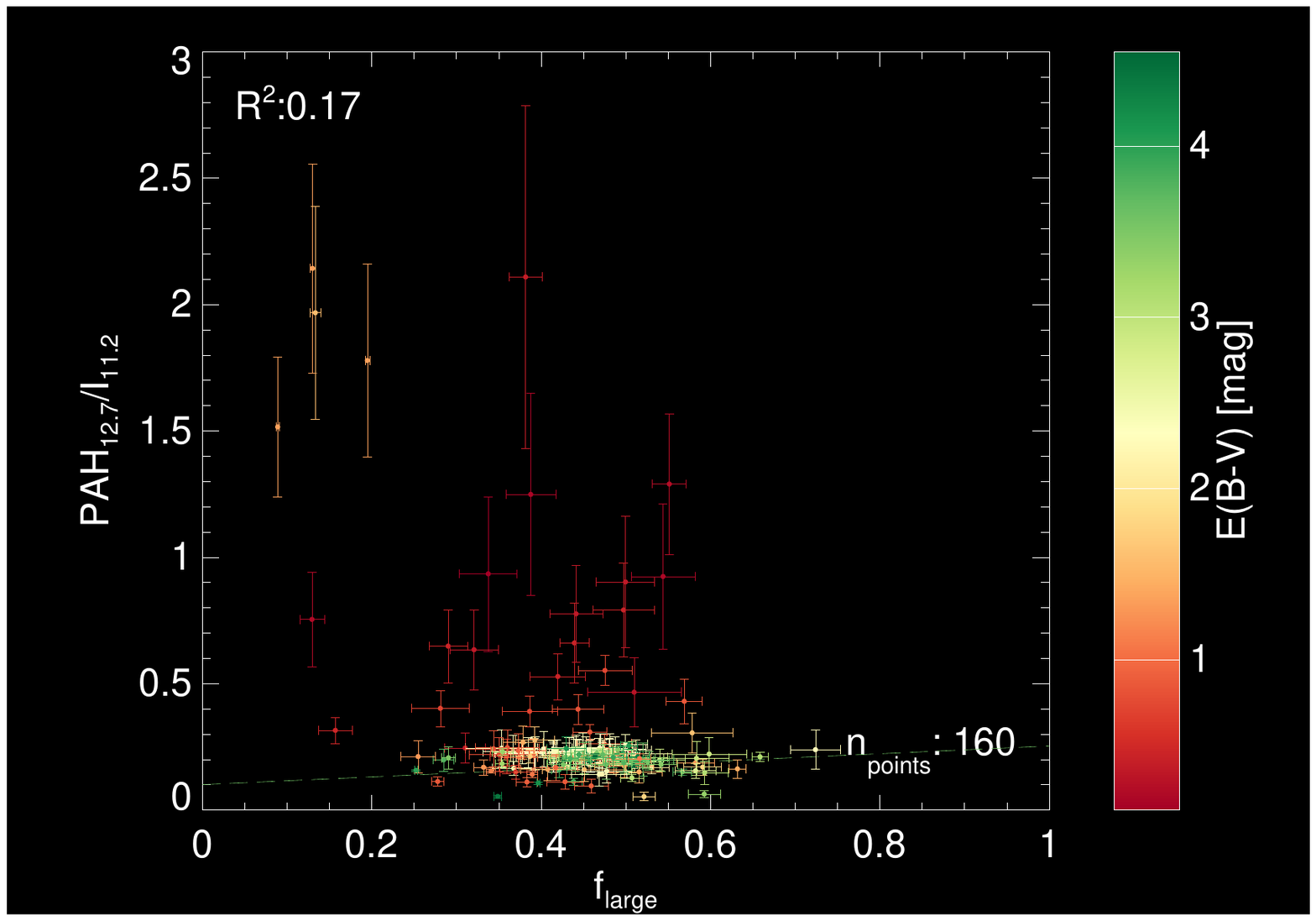}
  \caption{Top: The 6.2/11.2 (left) and 12.7/11.2~{\textmu}m (right) PAH band strength ratio versus the PAHdb-derived ionization fraction ($f_{\rm i}$). Middle: The 6.2/11.2 (left) and 12.7/11.2~{\textmu}m (right) PAH band strength ratio versus the PAHdb-derived large PAH fraction ($f_{\rm large}$). Bottom: The PAH$_{\rm 12.7}$/I$_{\rm 11.2}$~{\textmu}m PAH band strength ratio versus $f_{\rm i}$ (left) and $f_{\rm large}$ (right). The data have a SNR $\geq$ 3 and 0.06 $\leq$ E(B V) $\leq$ 5.0. Trend lines determined by fitting the data have been overlain. For each correlation the linear correlation coefficient R$^{\rm 2}$, that takes uncertainties into account, has been provided. NB Some extraneous data have been clipped for presentation purposes. See Sect.~\ref{subsec:fitting} for details.}
  \label{fig:parameters}
\end{figure*}

The top two panels in Fig.~\ref{fig:parameters} present the 6.2/11.2 (left panel) and 12.7/11.2~{\textmu}m (right panel) PAH band strength ratios plotted against the Monte Carlo-derived $f_{\rm i}$, while the two middle panels do the same for $f_{\rm large}$. The two bottom panels of Fig.~\ref{fig:parameters} take the PAH$_{\rm 12.7}$ PAH band strength instead of I$_{\rm 12.7}$. Trend lines have been added, constructed from straight-line fits to the data, and R$^{\rm 2}$ values that take uncertainties into account have been provided. Because of the 6.2~{\textmu}m PAH band falling in the poor-quality SL2 part of the spectrum, the number of points in the correlations involving it are sparse (64).

Notwithstanding the low number of 6.2/11.2~{\textmu}m data points, an interesting trend emerges from Fig.~\ref{fig:parameters} for the correlations involving the 12.7~{\textmu}m PAH band (as reflected by their R$^{\rm 2}$ values). While none of the correlations are particularly strong, the 6.2/11.2~{\textmu}m PAH band strength versus ionization fraction, $f_{\rm i}$, shows an overall positive trend while that of the 12.7/11.2~{\textmu}m PAH band strength versus $f_{\rm i}$ hints at a negative trend. This is somewhat surprising as a positive correlation between the 6.2/11.2 versus 12.7/11.2~{\textmu}m PAH band strength ratio has been well-established for ISM sources \citep[e.g.,][]{2001A&A...370.1030H, 2014ApJ...795..110B}. As shown in the lower two frames of the figure, this tentative negative trend also holds when considering PAH$_{\rm 12.7}$ instead of $I_{\rm 12.7}$, where R$^{\rm 2}$ improves somewhat. Turning to the large PAH fraction, $f_{\rm large}$, the negative trend with the 6.2/11.2~{\textmu}m PAH band strength ratio is again surprising as the correlation between $f_{\rm i}$ and $f_{\rm large}$ is generally shown to be positive for ISM sources \citep[e.g.,][]{2015ApJ...806..121B}. The same holds for the 12.7/11.2~{\textmu}m PAH band strength ratio. This unexpected behavior strongly suggests that PAH edge structure is an important \emph{variable} parameter that should be considered when analyzing (diffuse) ISM PAH background spectra, particularly when using ratios of individual band strengths as proxies for single PAH properties such as charge and size. As done here, individual bands are often normalized to the strong, well-defined 11.2~{\textmu}m band. However, the 11.2, 12.2, 12.7, 13.5, and 14.2~{\textmu}m PAH bands are produced by CH out-of-plane bending motions (CH$_{\rm oop}$), associated with solo, duo, trio, quartet, and quintet adjacent hydrogen atoms per edge ring, respectively \citep[e.g.,][]{2001A&A...370.1030H}. The number of these different hydrogen adjacency types depend on PAH structure and size, with small and irregularly shaped PAHs generally carrying substantially more duo through quintet hydrogens than solo hydrogens, while very large, compact structures with straight edges are dominated by solo hydrogens. The relative intensities of these bands vary with PAH size and structure. Thus, one could argue here that, when considering the 12.7/11.2~{\textmu}m PAH band strength ratio as a pure measure for PAH edge structure, as PAH size increases the relative number of solo hydrogens increases and the 11.2~{\textmu}m increases, causing the 12.7/11.2~{\textmu}m PAH band strength ratio to decrease.

Of course, the R$^{\rm 2}$ values are not particularly convincing, though some of that could be driven by the outliers. In addition, the results are likely susceptible to the systematic effect of the poor quality of much of the SL2 data. Nonetheless, as illustrated by Fig.~\ref{fig:pahdb_fit}, the fit to the 6.2~{\textmu}m band is somewhat constrained by the error bars. This points to the importance of having a complete $\sim$5-15~{\textmu}m spectrum when fitting \citep[see e.g.,][]{2015ApJ...806..121B}.

\subsection{PAH Spectra Averaged by E(B-V)}
\label{subsec:average}

\begin{figure*}
  \centering
  \includegraphics[width=\linewidth]{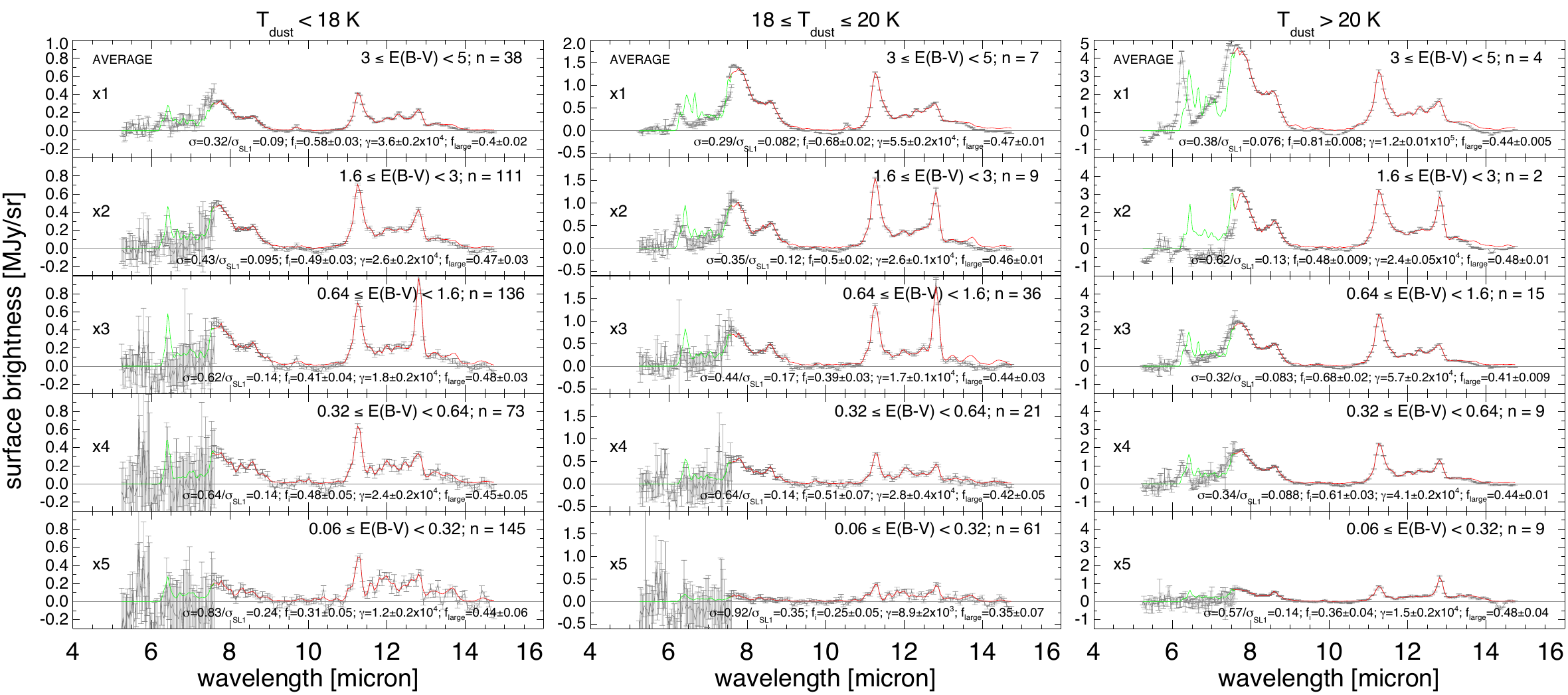}
  \caption{Average background PAH spectra across five E(B-V) and three T$_{\rm dust}$ bins. Only those background spectra with an uncertainty of less than 3$\times10^{\rm -21}$~W cm$^{\rm -2}$ for their 11.2~{\textmu}m PAH band strength and 0.06 $\leq$ E(B-V) $\leq$ 5 have been considered. The number of spectra in each bin is indicated by n. The average spectra (SL1+SL2 in grey) have been fitted (red=SL1, green=SL2) with PAH emission spectra synthesized using PAHdb. Indicated are the error ($\sigma$ and $\sigma_{\rm SL1}$), ionization- ($f_{\rm i}$) and large PAH-fraction ($f_{\rm large}$), and the PAH ionization parameter $\gamma$ determined from the fits through a Monte Carlo technique with the derived uncertainty given in parenthesis. See Sect.~\ref{subsec:average} for details.}
  \label{fig:averages}
\end{figure*}

To increase spectral fidelity, Fig.~\ref{fig:averages} takes the background SL1 and SL2 spectra (in grey) that have an uncertainty associated with the 11.2~{\textmu}m PAH band strength of less than 3$\times10^{\rm -21}$~W cm$^{\rm -2}$ and averages their broad band continuum subtracted spectra, normalized to the emission at 10~{\textmu}m, across five E(B-V) and three T$_{\rm dust}$ bins. The figure shows a steady increase in the overall PAH emission when moving both towards higher E(B-V) values and dust temperatures. This is accompanied by changes in the relative strengths of the 11.2 and 12.7~{\textmu}m PAH bands. Especially noticeable is the increased fidelity of the SL2 segment when moving towards higher E(B-V) and T$_{\rm dust}$ values.

The average PAH spectra are, as in Sect.~\ref{subsec:fitting}, fitted using PAHdb, and show generally good matches in the SL1 region of the spectra. However, the poor SNR at the shorter wavelengths covered by the SL2 segment are poorly constrained for all but the bins with the higher E(B-V) and T$_{\rm dust}$ values. Note that the SL3 bonus order has been discarded. The figure indicates both the single-run total error ($\sigma$) and that for the SL1 segment alone $\sigma_{\rm SL1}$, as well as $f_{\rm i}$ and $f_{\rm large}$ with their associated uncertainties derived from the fit using a Monte Carlo technique. $\sigma$ and $\sigma_{\rm SL1}$ systematically drop as E(B-V) increases. $f_{\rm i}$ generally increases when moving up an E(B-V) bin or crossing a T$_{\rm dust}$ boundary. Compared to $f_{\rm i}$, $f_{\rm large}$ shows less variance, typically staying between 0.4-0.5.

The almost systematic variations observed in $f_{\rm i}$, and perhaps $f_{\rm large}$ less so, imply altered PAH populations driven by changing astrophysical environments. This can be quantified by turning to the PAH ionization parameter $\gamma$, which relates the ionization state of the PAH population to the the strength of the radiation field (G$_{\rm 0}$), the electron density ($n_{\rm e}$), and the temperature of the gas (T$_{\rm gas}$) as $\gamma\propto$ G$_{\rm 0}$T$_{\rm gas}^{\rm 1/2}$/n$_{\rm e}$ \citep[see][]{2005pcim.book.....T}. Figure~\ref{fig:averages} provides the PAH ionization parameter $\gamma$ inferred from the fit as 2.66$\cdot f_{\rm i}$/$f_{\rm 0}$ [$\times10^{\rm 4}$ K$^{\rm 1/2}$ cm$^{\rm 4}$], where $f_{\rm 0}$ is the neutral PAH fraction \citep[e.g.,][]{2018ApJ...858...67B}. These are listed in Table~\ref{tab:gamma}.

\begin{deluxetable}{cccc}
 \tablecaption{PAHdb-derived PAH Ionization Parameter $\gamma$ for different E(B-V) and T$_{\rm dust}$ Bins. \label{tab:gamma}}
 \tablehead{
  \colhead{} &\multicolumn{3}{c}{$\gamma$ [$\times$10$^{\rm 4}$ K$^{\rm 1/2}$ cm$^{\rm 3}$]} \\
   \colhead{E(B-V)} & \colhead{T$_{\rm dust}$\textless18K} &
   \colhead{18$\leq$T$_{\rm dust}{\leq}$20K} &
   \colhead{T$_{\rm dust}$\textgreater20K}
 }
 \startdata
  0.06 - 0.32	& 1.2$\pm$0.2 & 0.89$\pm$0.2 & 1.5$\pm$0.2 \\
  0.32 - 0.64 & 2.4$\pm$0.2 & 2.8$\pm$0.4 & 4.1$\pm$0.2 \\
  0.64 - 1.6  & 1.8$\pm$0.2 & 1.7$\pm$0.1 & 5.7$\pm$0.2 \\
  1.6 - 3.0   & 2.6$\pm$0.2 & 2.6$\pm$0.1 & 2.4$\pm$0.05 \\
  3.0 - 5.0   & 3.6$\pm0.2$ & 5.5$\pm$0.2 & 12$\pm$0.1 \\
 \enddata
\end{deluxetable}

In general, but not across the board, the table shows the highest values of $\gamma$ for the warmest dust temperatures. The lower values of $\gamma$ are consistent with the warm neutral medium ($\gamma\simeq10^{4}$~K$^{\rm 1/2}$ cm$^{3}$), with the others pushing into the photo-dissociation region (PDR) domain ($\gamma\simeq$ few$\times$$10^{4}-10^{5}$~K$^{\rm 1/2}$ cm$^{3}$;  \citealt{2005pcim.book.....T}).

This shows that many of the background positions are not entirely isolated and are, to one degree or other, influenced by an additional radiation source rather than the interstellar radiation field alone. Obviously, the PDR-like backgrounds are far from isolated. By happenstance the off-target background position could have fallen on a nearby radiation source or it was simply not sufficiently separated from the extended on-source target. This could indeed be easily the case for a close-by star-forming region like the Orion Molecular Cloud.

\section{Astronomical Implications}
\label{sec:implications}

A sizable fraction (18\%; SNR$_{\rm 11.2}>$ 3) of the background spectra show detectable PAH emission, with the bulk (85\%) in directions where 0.06 $\leq$ E(B-V) $\leq$ 5.0.

While the correlation between PAHs and classical dust shown in Fig.~\ref{fig:correlations} appears at first glance to be poor, with a wide range in PAH emission strengths at all values of E(B-V), the correlations of the 11.2 and 12.7~{\textmu}m PAH bands with WISE 12 Band 3 observations point to a much stronger connection. Figure~\ref{fig:linear} shows these bands plotted linearly against E(B-V). As discussed in Sect.~\ref{subsec:extinction}, these plots show temperature stratification and a linear upper bound relationship between the PAHs and classical dust.

All of the points along the upper boundary have high dust temperatures while points with lower temperature dust fills in below the upper bound. This implies that PAH abundances and dust densities are well-correlated and that more intense radiation fields produce more PAH emission and warmer dust. This suggests that the PAH ionization parameter $\gamma$, which is connected to the intensity of the radiation field through G$_{\rm 0}$, would correlate with dust temperature when assuming the electron density n$_{\rm e}$ and gas temperature T$_{\rm gas}$ are largely invariant. Though, while Table~\ref{tab:gamma} indeed has the largest $\gamma$'s associated with the highest T$_{\rm dust}$ bins, a clear one-to-one correlation is lacking. Inspection of Fig.~\ref{fig:averages} shows signs of Ne~II 12.8~{\textmu}m emission in some of the averaged spectra (See also Sect.~\ref{subsec:lines}). This indicates possible contributions from ionized regions along those lines of sight and hence those values of $\gamma$ may not be indicative of the isolated background ISM.

Each line of sight is a composite of all the material along each direction and consists of regions with differing densities and illumination. Those lines of sight with the lowest E(B-V) values can only be composed of low extinction regions, while those with high E(B-V) values could include a multitude of low, moderate, and high extinction regions. The lowest extinction lines of sight have low $\gamma$s, indicative of the warm neutral phase of the ISM, while the other lines of sight are consistent with PAH emission from PDRs. In this case a PDR includes any cloud or filament surface with somewhat higher density than the warm neutral phase that is excited by non-ionizing UV photons. Those directions that show obvious Ne~II line emission must have a component that is illuminated by ionizing UV photons from luminous O and B stars.

One last thing to consider is the separation of the background from the intended science target and, on a larger scale, that from associated cloud structures. The former seems to be sufficiently scrutinized by the cross-dispersion profiles of the 11.2~{\textmu}m PAH emission, which are spatially completely unresolved for \emph{all} backgrounds (Fig.~\ref{fig:dispersion}).

Concerning the latter, the presence of Ne~II line emission in at least some of the background spectra does seem to indicate, not unexpectedly, that some of the backgrounds are associated with luminous O-B stars likely connected to larger extended structures. Appendix.~\ref{app:clustering} examines this in more detail for the more-diffuse backgrounds (0.06 $\leq$ E(B-V) $\leq$ 5.0) by first spatially clustering by position and then constructing images of the 128 cluster regions from GAIA and WISE 12 full-sky dust maps. Indeed, many regions show the background positions to be associated with filamentry cloud-like structures.

\section{The Ames Background Interstellar Medium Spectral Catalog}
\label{sec:catalogue}

The Ames Background Interstellar Medium Spectral Catalog makes available the non-dark off-module extracted background spectra as well as a separate downloadable complementary table containing all derived measurement from this work. The catalog can be accessed at \url{www.astrochemistry.org/bism}. Spectra can be retrieved using a known AOR-key, by specifying coordinates, or by providing a target name that will be resolved using SIMBAD services. In the latter two cases the catalog is scanned at 1$^{\circ}$ increments from the resolved position until there is at least a single hit. Figure~\ref{fig:landing} shows the search interface as presented at the website.

\newlength{\bwidth}
\setlength{\bwidth}{\dimexpr(\linewidth-12pt)\relax}

\begin{figure}
  \centering
  \shadowpicture{\includegraphics[width=\bwidth]{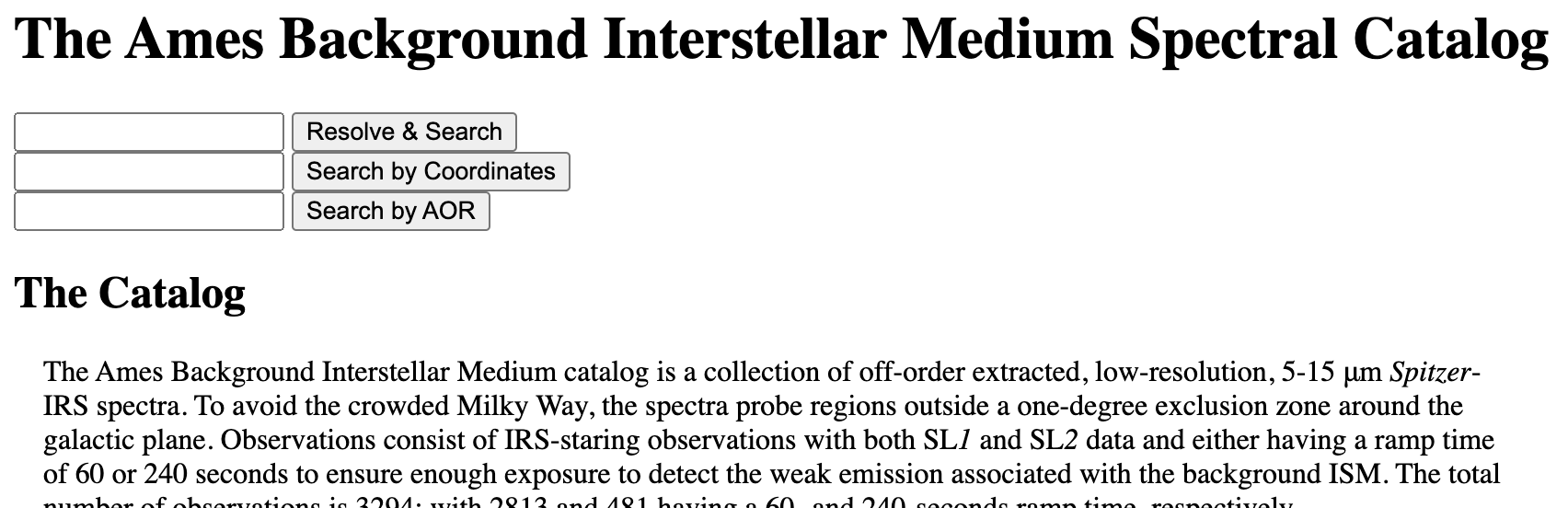}}
  \caption{The Ames Background Interstellar Medium Spectral Catalog's landing page showing the search interface.}
  \label{fig:landing}
\end{figure}

\begin{figure}
  \centering
  \shadowpicture{\includegraphics[width=\bwidth]{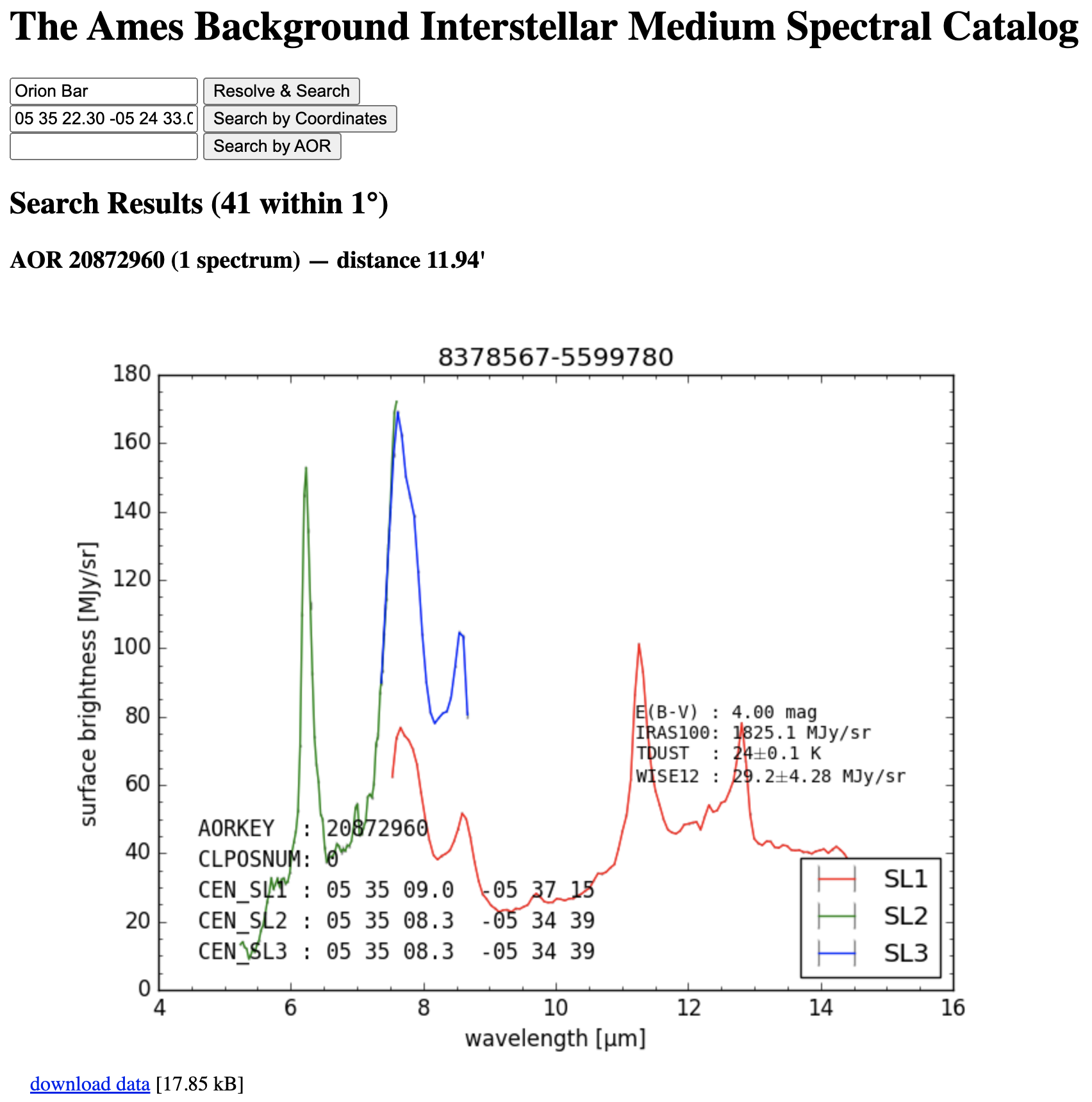}}
  \caption{The Ames Background Interstellar Medium Spectral Catalog's results page showing one of the non-dark subtracted spectra found when searching for `Orion Bar'.}
  \label{fig:result}
\end{figure}

 Query results are organized per AOR-key and the number of available spectra are indicated. For each result a spectrum is shown, with the orders color-coded separately and associated error bars. Also provided are the statistical representations for E(B-V), IRAS100~{\textmu}m emission, the dust temperature as retrieved from the Galactic Dust Reddening and Extinction service at IPAC, and WISE12 measurements determined as described in this work. The spectra can be downloaded in IPAC format by clicking the `download data'-link. The file size is indicated for convenience. Figure~\ref{fig:result} shows an example when querying and resolving for `Orion Bar'.

\begin{figure}
  \centering
  \shadowpicture{\includegraphics[width=\bwidth]{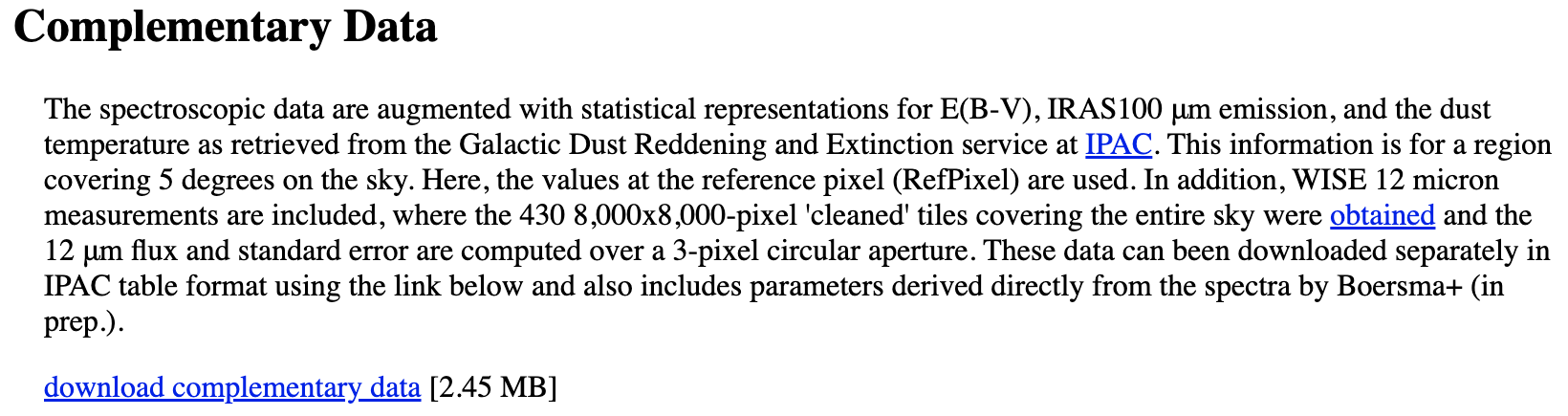}}
  \caption{Ames Background Interstellar Medium Spectral Catalog website showing the link to the complementary data.}
  \label{fig:aux}
\end{figure}

 The link to download the complementary data can be found near the bottom of the page, as is shown in Fig.~\ref{fig:aux}.

The website is written in PHP8 with SQLite3 as the database back-end and runs on Apache2 under Ubuntu Server 22.04 LTS. The figures and downloadable data have all been pre-produced.

\section{Summary and Conclusions}
\label{sec:conclusions}

Many data sets obtained by the Spitzer-IRS contain `hidden' observations of the IR background ISM. These are often used to subtract line-of-sight contamination, or are simply ignored. Here, these background observations are considered in their own right. A catalog of 4,090 spectra is constructed to examine the PAH spectral signature in the background ISM and its connection to extinction by (classical) dust.

To isolate any background PAH emission, carefully selected high-galactic latitude spectra, dominated by zodiacal light, were averaged and subtracted from the background data. Subsequently, the 6.2, 11.2, and 12.7~{\textmu}m PAH band strengths were determined. \textit{A strong, positive correlation is recovered between the 11.2~{\textmu}m PAH band strength with extinction (E(B-V)) and WISE12 observations}.

Focusing on the more-diffuse emission with 0.06 $\le$ E(B-V) $\le$ 5.0, correlations of the 6.2, 11.2, and 12.7~{\textmu}m PAH band strength is positive with E(B-V), albeit with considerable scatter. In addition, the correlations reveal a clear separation into three distinct dust temperature regimes when presented on a linear scale. The correlations with WISE12 data are far better constrained.

Decomposition of the PAH emission in terms of charge and size using the data and tools made available through PAHdb reveals a tentative positive correlation between the 6.2/11.2~{\textmu}m PAH band strength ratio and $f_{\rm i}$, while that with the 12.7/11.2~{\textmu}m PAH band strength ratio, surprisingly, hints to a slight negative trend. The 12.7/11.2~{\textmu}m PAH band strength ratio normally tracks with PAH ionization along with the 6.2/11.2~{\textmu}m PAH band strength ratio. Since the 11.2 and 12.7~{\textmu}m bands are also tracers for PAH edge structure and size, this behavior suggests PAH structures are changing along these lines of sight.

The relation between $f_{\rm large}$ and the 6.2/11.2~{\textmu}m PAH band strength ratio is negative, with that with the 12.7/11.2~{\textmu}m PAH band strength ratio hinting at being positive.

Increasing the SNR by averaging the background spectra into five E(B-V) and three T$_{\rm dust}$ bins shows a clear evolution in the strength of the PAH emission and variations in the relative strength of the 11.2 and 12.7~{\textmu}m PAH bands. Database-fits show, overall, an increase in $f_{\rm i}$ and $\gamma$ but a somewhat more stable $f_{\rm large}$. While the largest found $\gamma$s are associated with the highest T$_{\rm dust}$ bins, a clear one-to-one correlation is lacking. However, much of the analysis remains, in many cases, limited by the low SNR at shorter wavelengths ($\lambda\lesssim$ 7.5~{\textmu}m).

Taking everything together, there are some hints that the PAH population in the more-diffuse background behaves differently from that of the general ISM. However, in most cases the backgrounds are still associated with larger scale filamentary cloud-like structures or, in a few cases, PDR-like environments located somewhere along the line of sight.

The spectra and auxiliary data have been made publicly available for download through the Ames Background Interstellar Medium Spectral Catalog and, as they may guide JWST programs, focused on explicitly studying the (more-diffuse) background ISM.

\begin{acknowledgments}

 C.B. is grateful for an appointment at NASA Ames Research Center through the San Jos\'e State University Research Foundation (80NSSC22M0107). C.B., J.D.B., L.J.A. and A.M. acknowledge support from the Internal Scientist Funding Model (ISFM) Laboratory Astrophysics Directed Work Package at NASA Ames (22-A22ISFM-0009). L.J.A., J.D.B. and A.M. are thankful for an appointment at NASA Ames Research Center through the Bay Area Environmental Research Institute (80NSSC19M0193).

\end{acknowledgments}

\vspace{5mm}
\facilities{Spitzer}

\software{astropy \citep[][]{2013A&A...558A..33A,2018AJ....156..123A},
     amespahdbidlsuite \citep[][]{2018ApJS..234...32B}}

\bibliography{aamnem99,bibliography}{}
\bibliographystyle{aasjournal}

\appendix

\setcounter{table}{0}
\renewcommand{\thetable}{A\arabic{table}}

\section{Observations with Complications}
\label{app:complications}

Table~\ref{app:tab:complications} lists and provides a brief description of those observations for which analysis complications were encountered.

\begin{deluxetable}{ll}
 \tablecaption{Observations with Complications. \label{app:tab:complications}}
 \tablehead{
  \colhead{AOR key} & \colhead{Description}}
 \startdata
 14136064 & no BCD files \\
 23039488 & no BCD files \\
 26086912 & CUBISM `no fluxcon ...' error (for NODARK) \\
 23796224 & CUBISM `no fluxcon ...' error (for NODARK) \\
 23795968 & NULL-pointer reference in CUBEPROJ::BUILDCUBE 4265 (for NODARK) \\
 \enddata
\end{deluxetable}




\section{Spatial Clustering}
\label{app:clustering}

\begin{figure}
  \centering
  \includegraphics[width=\linewidth]{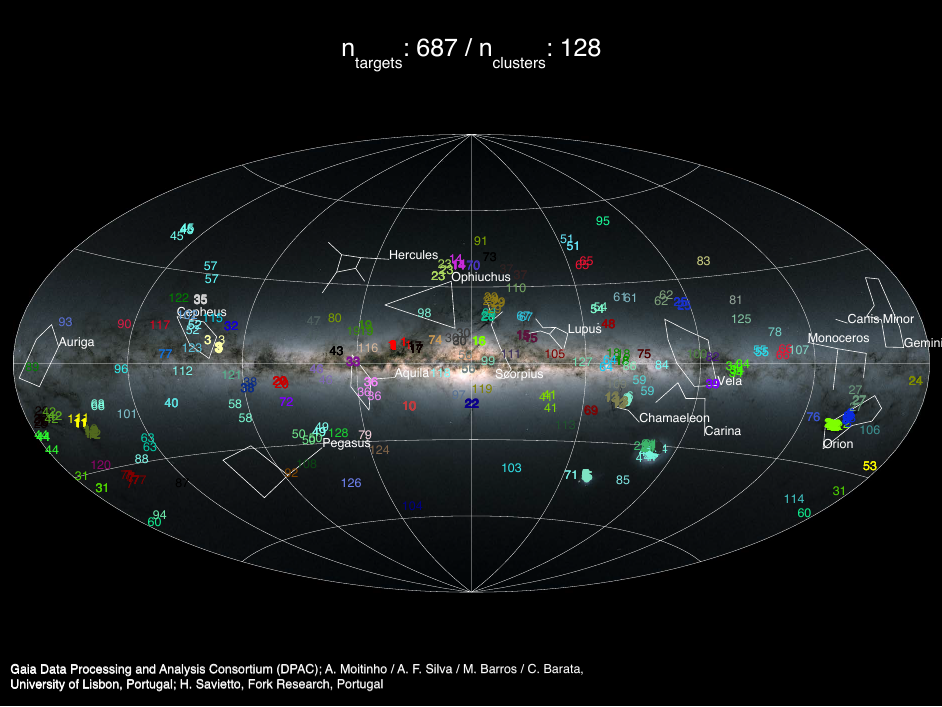}
  \caption{Position of the 687 backgrounds with an associated uncertainty of less than 3$\times10^{\rm -21}$~W cm$^{\rm -2}$ for the 11.2~{\textmu}m PAH band strength and 0.06 $\leq$ E(B-V) $\leq$ 5.0. Each background position is indicated by its region number established through hierarchical clustering and uniquely color-coded.}
  \label{fig:clusters}
\end{figure}

\begin{figure*}
  \centering
  \includegraphics[width=\linewidth]{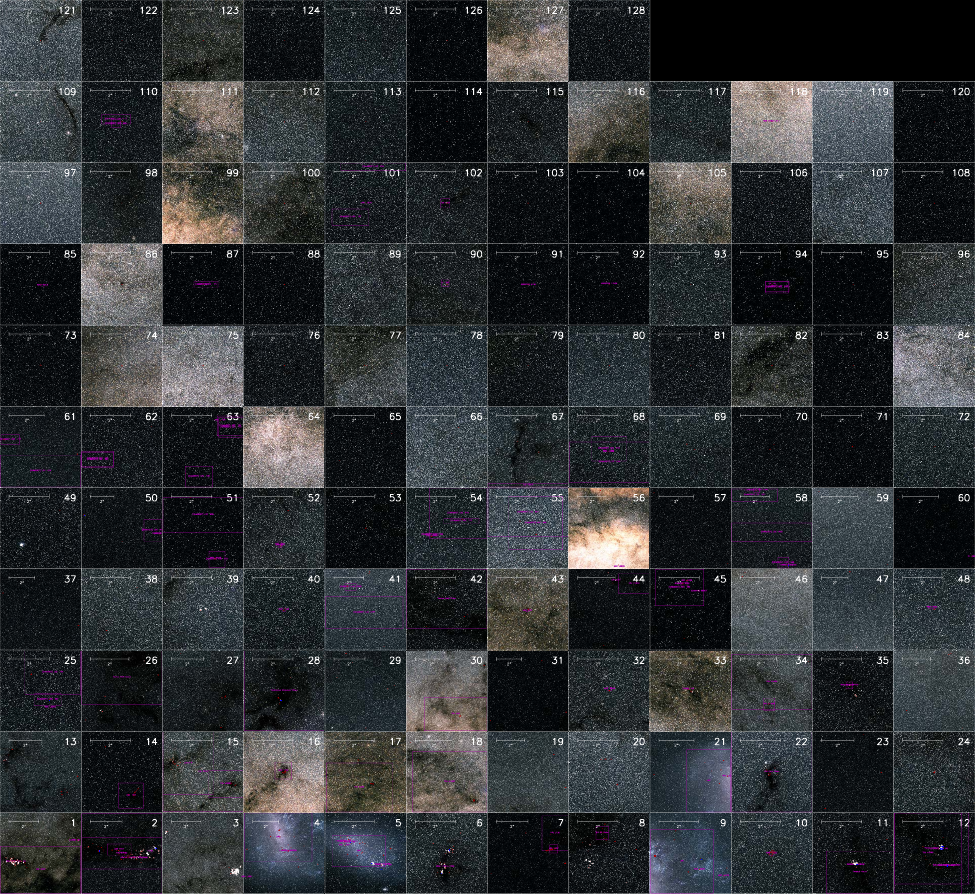}
  \caption{Zoom-in on the 128 cluster regions from Fig.~\ref{fig:gaia_stamps} using the 40kx20k GAIA all-sky image. Scale bars indicate 2$^{\circ}$. In white nearby SIMBAD IR sources are shown and in blue those with known sizes. Parent objects are indicated in purple, clipped to the extent of the field-of-view. Acknowledgment: Gaia Data Processing and Analysis Consortium (DPAC); A. Moitinho / A. F. Silva / M. Barros / C. Barata, University of Lisbon, Portugal; H. Savietto, Fork Research, Portugal.}
  \label{fig:gaia_stamps}
\end{figure*}

\begin{figure*}
  \centering
  \includegraphics[width=\linewidth]{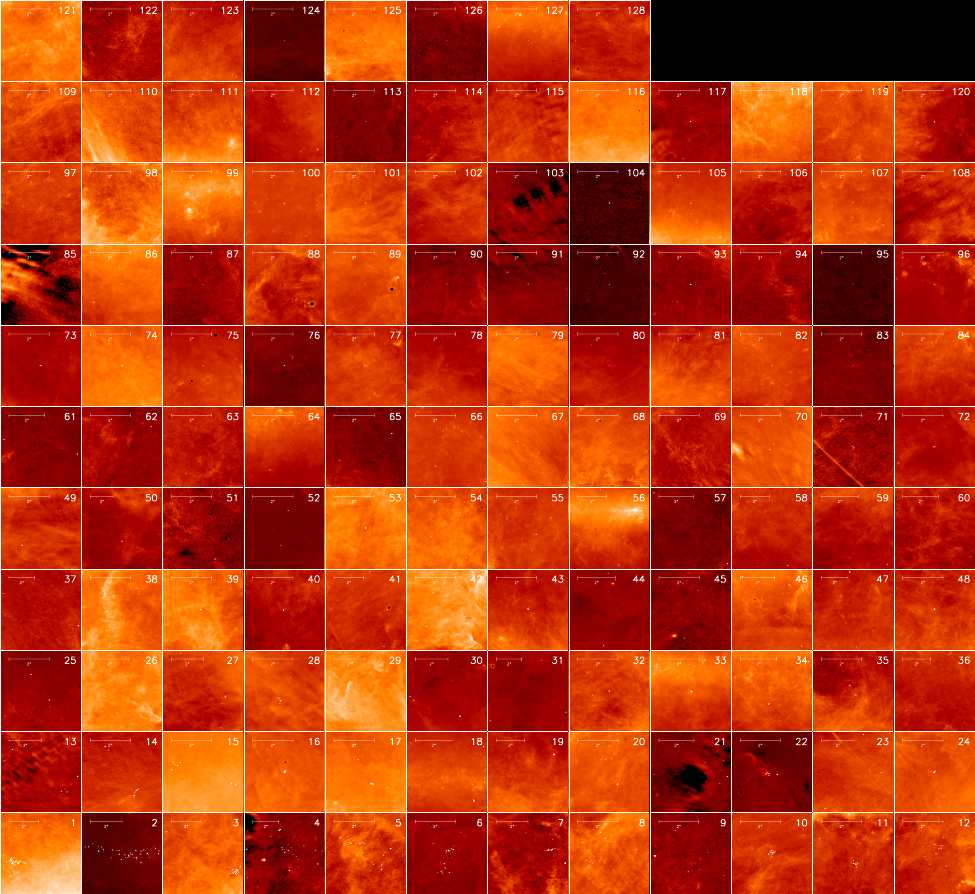}
  \caption{Zoom-in on the 128 cluster regions from Fig.~\ref{fig:clusters} using the 430, 8,000x8,000-pixel `cleaned' tiles from the WISE 12 micron full-sky dust map. Scale bars indicate 2$^{\circ}$. The background positions are shown as the white points. Each image is displayed using a logarithmic scaling.}
  \label{fig:WISE_stamps}
\end{figure*}

Figure~\ref{fig:clusters} shows the background positions where the uncertainty associated with the 11.2~{\textmu}m PAH band strength is less than $3\times10^{-21}$~W cm$^{\rm -2}$ and 0.06 $\leq$ E(B-V) $\leq$ 5.0. The positions have been grouped into regions using hierarchical clustering based on complete linkage and a maximum link distance of 6$^\circ$. Figure~\ref{fig:gaia_stamps} zooms in on each of the regions using the 40kx20k-pixels all-sky GAIA color image (2$^{\rm nd}$ data release) and indicates each background position by its cluster number and a unique color. Utilizing SIMBAD's TAP service\footnote{\url{simbad.u-strasbg.fr}}, IR sources within a 117" radius of each position were identified as well as any hierarchical links. This information is displayed in Fig.~\ref{fig:WISE_stamps}, where IR sources are shown as white dots or as blue boxes when size information is available. Parent objects are indicated in purple. Many regions show the background positions associated with dark filamentary-like cloud structures.

Figure~\ref{fig:WISE_stamps} does the same as Fig.~\ref{fig:gaia_stamps}, but now uses the 430, 8,000x8,000-pixel WISE 12 micron full-sky dust map. For each region the appropriate sub-images were extracted from the contributing `clean' WISE-tiles and combined into a single image. The images are displayed using a logarithmic scaling.


\end{document}